\documentclass{elsarticle}
\usepackage[margin=1in]{geometry} 

\usepackage{amssymb}
\usepackage{amsmath}
\usepackage[colorlinks]{hyperref}
\usepackage{cleveref}
\usepackage{siunitx}
\usepackage{graphicx}
\usepackage{xspace}
\usepackage{listings}
\usepackage{inconsolata}
\usepackage{fancyvrb}
\usepackage{color}
\crefname{lstlisting}{listing}{listings}
\Crefname{lstlisting}{Listing}{Listings}

\newcommand{\code}[1]{\lstinline|#1|}
\newcommand{\addpp}[1]{{#1\nolinebreak[4]\hspace{-.05em}\raisebox{.4ex}{\tiny\bf ++}}\xspace}
\newcommand{\CXX}{\addpp{C}}
\newcommand{\www}[1]{\href{#1}{#1}}
\newcommand{\diag}{\mathop{\mathrm{diag}}}

\def\lang#1{#1}

\journal{Journal of Computational Science}

\begin{document}
\begin{frontmatter}

\title{Accelerating linear solvers for Stokes problems\\with C++ metaprogramming}

\author{Denis Demidov\corref{c1}\fnref{f1}}
\cortext[c1]{Corresponding author}
\ead{dennis.demidov@gmail.com}
\address{
    Kazan Branch of Joint Supercomputer Center, Scientific Research Institute of System Analysis,\\
    the Russian Academy of Sciences, Lobachevsky st. 2/31, 420111 Kazan, Russian Federation}

\author{Lin Mu}
\ead{linmu@uga.edu}
\address{Department of Mathematics, University of Georgia, Athens, GA, 30605, USA}

\author{Bin Wang}
\ead{binwang.0213@gmail.com}
\address{
    Craft \& Hawkins Department of Petroleum Engineering
    Louisiana State University, Baton Rouge, LA  70803, USA}

\fntext[f1]{The development of the AMGCL library was partially funded by the
    state assignment to the Joint supercomputer center of the Russian Academy
    of sciences for scientific research. Work on the Schur pressure correction
    preconditioner was partially funded by the RFBR grant Nos 18-07-00964,
    18-47-160010.}

\begin{abstract}
    The efficient solution of large sparse saddle point systems is very
    important in computational fluid mechanics. The discontinuous Galerkin
    finite element methods have become increasingly popular for incompressible
    flow problems but their application is limited due to high computational
    cost. We describe \CXX programming techniques that may help to
    accelerate linear solvers for such problems. The approach is based on the
    policy-based design pattern and partial template specialization, and is
    implemented in the open source AMGCL library. The efficiency is
    demonstrated with the example of accelerating an iterative
    solver of a discontinuous Galerkin finite element method for the Stokes
    problem. The implementation allows selecting algorithmic components of the solver by
    adjusting template parameters without any changes to the codebase.  It
    is possible to switch \lang{the system matrix to use small statically sized
    blocks to store the nonzero values}, or use a mixed precision
    solution, which results in up to 4 times speedup, and reduces the memory
    footprint of the algorithm by about 40\%. We evaluate both monolithic and
    composite preconditioning strategies for 3 benchmark problems. The
    performance of the proposed solution is compared with a multithreaded direct
    Pardiso solver and a parallel iterative PETSc solver.
\end{abstract}


\end{frontmatter}

\section{Introduction}

Incompressible flow modeling has many applications in the scientific and
engineering field \cite{golparvar2018comprehensive, sun20206m}. The discontinuous
Galerkin (DG) finite element methods (FEM) have become increasingly popular for
incompressible flow problems (Stokes or \lang{Navier-Stokes} equations). Compared to
the continuous Galerkin (CG) FEM, DG offers several
attractive features for incompressible flow simulations: higher-order accuracy
on arbitrary unstructured meshes, local mass and momentum conservation,
pointwise divergence-free velocity solution, and ease of parallelization
\cite{cockburn2012discontinuous,Sime2020AnE,landmann2008parallel}. However, the
high computational cost of DG methods limits their applicability.  Recently,
several hybridized DG methods (hybridizable discontinuous Galerkin and weak
Galerkin methods) have been proposed to reduce the global degrees-of-freedom
(DOFs) of DG while retaining many of the attractive features of DG methods
\cite{maljaars2020leopart,bin2020weak,mu2020pressure}. In these hybridized DG
methods, the number of global DOFs is significantly reduced by introducing facet
variables and eliminating local DOFs via static condensation. However, to
achieve the same level of accuracy, the DOFs and computational cost of the
hybridized DG methods are still much higher than the conventional CG FEM
\cite{gibson2020slate,yakovlev2016cg}. 

For a large 3D simulation, both memory footprint and the solution
performance depends not only on the problem size but also on the linear solver
configuration. For the Stokes problems, the discretization of the conventional
CG FEM results in an indefinite saddle point linear system. There is a lot of
research dedicated to the effective solution of such systems,
see~\cite{benzi2005numerical} for an extensive overview. Common approaches are
direct solvers and preconditioned Krylov subspace iterative solvers. The direct
methods are robust, but do not scale beyond a certain size, typically of the
order of a few millions of unknowns~\cite{hogg2013new,henon2002pastix}, due to
their intrinsic memory requirements and sheer computational cost.
Preconditioned iterative solvers may be further \lang{classified} by the type of the
preconditioner. The simplest choice would be a monolithic preconditioner that
does not take the block structure of the linear system into consideration and
treats the problem as an opaque linear system. This approach usually does not
work very well for saddle point problems, but may still be considered as a
viable solution in some cases.  Another large class of preconditioners for
saddle point problems takes the block structure of the system matrix into
account. Well-known examples of such preconditioners are inexact Uzawa
algorithm~\cite{elman1994inexact}, SIMPLE schemes~\cite{patankar1980numerical},
or block-triangular preconditioners~\cite{bramble1988preconditioning}. Here we
are using the Schur complement pressure correction
preconditioner~\cite{verfurth1984combined, saleri2005pressure,
gmeiner2016quantitative}. To our best knowledge,
the applicability and performance of the various iterative solvers for the
Stokes problem that are commonly used by the CG FEM remains unknown for the new
hybridized DG methods.

A lot of popular scientific software packages today are either developed with C
or Fortran programming languages, or have C-compatible application programming
interface (API). Notable examples are PETSc~\cite{petsc-user-ref} software
package, or BLAS and LAPACK standard programming interfaces with well known and
efficient implementations such as Intel MKL~\cite{wang2014intel},
OpenBLAS~\cite{xianyi2012openblas}, or NVIDIA CUBLAS~\cite{nvidia2019cublas}.
The low-level API makes it easy to use the functionality, as most modern
programming languages support interaction with external libraries with a
C-compatible API. However, this also has some disadvantages: such packages
usually have fixed interfaces and may only deal with the cases that the
developers have thought about in advance. For example, BLAS has separate sets
of similar functions that deal with single, double, complex, or double complex
values, but it is impossible to work with mixed precision inputs or with
user-defined custom types despite some previous efforts~\cite{li2002design}.
The PETSc framework, despite being extremely configurable and flexible, still
does not support mixed precision solvers or non-standard value types, because
switching a value type is done with a preprocessor definition and has a global
effect.

In this work, we consider the \CXX programming techniques that support creating
flexible, extensible and efficient scientific software on the example of the
open source AMGCL\footnote{Published at \www{https://github.com/ddemidov/amgcl}
under MIT license.} library~\cite{Demidov2019, Demidov2020} that implements the
algebraic multigrid method (AMG)~\cite{brandt1985algebraic, Stuben1999} and
other preconditioners for solution of large sparse linear systems of equations.
The library uses the \CXX metaprogramming so that its users may easily extend it or
use it with their datatypes.  The advantages of this approach are studied
with an example of accelerating an iterative solver of a DG method,
specifically the weak Galerkin method, for the Stokes problem.  We show that static polymorphism
allows reusing the same code in order to exploit the block structure of the
matrix by switching to small statically sized matrices as value type, and to
employ the mixed precision approach. This results in up to 4 times faster
solution and up to 40\% reduction of the memory footprint of the algorithm.
The source code for the benchmarks is published \lang{in}~\cite{cppstokesbm},
and the dataset used in the benchmarks is made available \lang{in}~\cite{cppstokesds}.

The rest of the paper is organized as follows.  In \Cref{sec:amgcl} we discuss
the design choices behind the AMGCL library that improve the flexibility and
performance of the code. \Cref{sec:stokes} describes the Stokes problem and
weak Galerkin discretization. \Cref{sec:solvers} provides an overview of
the linear solvers used in this work. \Cref{sec:performance} presents the
results of the numerical experiments and compares the performance of various
solution approaches implemented in AMGCL with the performance of a direct MKL
solver and an iterative PETSc solver.

\section{AMGCL} \label{sec:amgcl}

AMGCL is a \CXX header-only template library with a minimal set of
dependencies and provides both shared-memory and distributed
memory (MPI) versions of the algorithms. The multigrid hierarchy is constructed
using builtin data structures and then transferred into one of the provided
backends. This allows for transparent acceleration of the solution phase with
help of OpenMP, OpenCL, or CUDA technologies. Users may even provide their
backends \lang{to enable a} tight integration between AMGCL and the user code.  The
library uses the following design principles:

\begin{itemize}
    \item \emph{Policy-based design}~\cite{alexandrescu2001modern} of public
        library classes such as \code{amgcl::make_solver} or \code{amgcl::amg}
        allows the library users to compose their customized version of the
        iterative solver and preconditioner from the provided components and
        easily extend and customize the library by providing their
        implementation of the algorithms.
    \item Preference for \emph{free functions} as opposed to member
        functions~\cite{meyers2005effective}, combined with \emph{partial
        template specialization} allows extending the library operations onto
        user-defined datatypes and to introduce new algorithmic components when
        required.
    \item The \emph{backend} system of the library allows expressing the
        algorithms such as Krylov iterative solvers or multigrid relaxation
        methods in terms of generic parallel primitives \lang{to facilitate}
        a transparent acceleration of the solution phase with OpenMP, OpenCL, or
        CUDA technologies.
    \item One level below the backends are \emph{value types}: AMGCL supports
        systems with scalar, complex, or block value types both in single and
        double precision. Arithmetic operations necessary for the library implementation
        may also be extended onto the user-defined types using template
        specialization.
\end{itemize}

\subsection{Policy-based design}

\begin{lstlisting}[float=t,
    caption={Policy-based design illustration: creating customized solvers from AMGCL components},
    label=lst:composition]
// GMRES solver preconditioned with AMG
typedef amgcl::make_solver<
  amgcl::amg<
    amgcl::backend::builtin<double>,
    amgcl::coarsening::smoothed_aggregation,
    amgcl::relaxation::spai0
    >,
  amgcl::solver::gmres<
    amgcl::backend::builtin<double>
    >
  > Solver;
\end{lstlisting}

The solvers and preconditioners in AMGCL are composed by the library
user from the provided components. For example, the most frequently used class
template \code{amgcl::make_solver<P,S>} binds an iterative solver \code{S}
to a preconditioner \code{P} chosen by the user. To illustrate this,
\Cref{lst:composition} defines a GMRES iterative solver preconditioned with the
algebraic multigrid (AMG). Smoothed aggregation is used as the AMG coarsening
strategy, and diagonal sparse approximate inverse (SPAI0) is used on each level
of the multigrid hierarchy as the smoother. Similar to the solver and the
preconditioner, the AMG components (coarsening and relaxation) are specified as
template parameters and may be customized by the user.  Here the double
precision builtin backend (parallelized with OpenMP) is used both for the
solver and the preconditioner. Choosing another backend like
\code{amgcl::backend::cuda<double>} or \code{amgcl::backend::vexcl<double>}
would make the solver use the CUDA or OpenCL for an acceleration of the solution.
This approach allows the library users not only to select any of the
solvers, preconditioners, or backends provided by AMGCL, but also to use their custom
components, as long as they conform to the generic AMGCL interface.

\begin{lstlisting}[float=t,
    caption={Example of parameter declaration in AMGCL components},
    label=lst:declparams]
template <class Precond, class Solver>
struct make_solver {
  struct params {
    typename Precond::params precond;
    typename Solver::params solver;
  };
};
\end{lstlisting}

Besides compile-time composition of the AMGCL algorithms described above,
it may be necessary to specify runtime parameters for the constructed
solvers.  This is done with the \code{params} structure declared by each of
the components as a subtype. In general, each parameter has a
reasonable default value. When a class is composed of several components, it
includes the parameters of its dependencies into its own \code{params} struct.
This allows providing a unified interface to the parameters for various AMGCL
algorithms. \Cref{lst:declparams} shows that the parameters for the
\code{amgcl::make_solver<P,S>} class are declared simply as a combination of the
preconditioner and the solver parameters. \Cref{lst:params} shows an example
of parameters specification for the solver from~\Cref{lst:composition}.
Namely, the number of the GMRES iterations before a restart is
set to 50, the relative residual threshold is set to $10^{-6}$, and the strong
connectivity threshold $\varepsilon_{str}$ for the smoothed aggregation is set
to $10^{-3}$.  The rest of the parameters are left with their default values.

\begin{lstlisting}[float=t,
    caption={Setting parameters for AMGCL components},
    label=lst:params]
// Set the solver parameters
Solver::params prm;
prm.solver.M = 50;
prm.solver.tol = 1e-6;
prm.precond.coarsening.aggr.eps_strong = 1e-3;

// Instantiate the solver
Solver S(A, prm);
\end{lstlisting}

\subsection{Free functions and partial template specialization}

Using free functions as opposed to class methods allows us to decouple the library
functionality from specific classes and enables support for third-party
datatypes within the library~\cite{meyers2005effective}. Delegating the
implementation of a free function to a struct template specialization
provides more control over the mapping between the input datatype and the
specific version of the algorithm.  For example, class constructors in AMGCL
may accept an arbitrary datatype as an input matrix, as long as the
implementation of specific functions supporting the datatype has been
provided. Some of the free functions that need to be implemented are
\code{amgcl::backend::rows(A)} and \code{amgcl::backend::cols(A)} (returning the
number of rows and columns for the matrix), or
\code{amgcl::backend::row_begin(A,i)} (returning iterator over the nonzero
entries for the matrix row). AMGCL provides adapters for several common input
matrix formats, such as \code{Eigen::SparseMatrix} from
Eigen~\cite{guennebaud2010eigen}, \code{Epetra_CrsMatrix} from Trilinos
Epetra~\cite{heroux2005overview}, or a generic tuple of the \lang{Compressed Sparse Row (CSR)} arrays.
It is also possible to adapt a user-defined
datatype.

\subsection{Backends}

A backend in AMGCL is a class binding datatypes like matrix and vector to
parallel primitives like matrix-vector product, linear combination of vectors,
or inner product computation. The backend system is implemented using the free
functions combined with a template specialization approach from the previous
section, which decouples the implementation of common parallel primitives from
specific datatypes used in the supported backends.
For example, in order to switch to the CUDA backend in
\Cref{lst:composition}, one just needs to replace
\code{amgcl::backend::builtin<double>} with
\code{amgcl::backend::cuda<double>}.
The backend system allows non-intrusive adoption of third-party or
user-defined datatypes for use within AMGCL.

An algorithm setup in AMGCL is performed using internal data structures. As soon
as the setup is completed, the necessary objects (mostly matrices and vectors)
are moved to the backend datatypes. The solution phase of an algorithm is
expressed in terms of the predefined parallel primitives which makes it
possible to select a parallelization technology (such as OpenMP, CUDA, or OpenCL)
simply by changing the backend template parameter of the algorithm. For
example, the \lang{L2 norm of the residual} $\epsilon = ||f - Ax||$ in AMGCL is computed using
\code{amgcl::backend::residual()} and \code{amgcl::backend::inner_product()}
primitives:

\begin{lstlisting}
backend::residual(f, A, x, r);
auto e = sqrt(backend::inner_product(r, r));
\end{lstlisting}

\subsection{Value types}

The value type concept allows generalizing AMGCL algorithms for complex or
non-scalar systems. A value type defines several overloads for common math
operations and is specified as a template parameter for a backend. Most often, a
value type is simply a builtin \code{double} or \code{float} scalar, but
it is also possible to use small statically sized matrices when the system
matrix has a block structure, which decreases the setup time and
memory footprint of the algorithm, increases cache locality and may improve convergence
rate in some cases~\cite{gupta2010adaptive}.

The value types are used during both the setup and the solution phases. Common
value type operations are defined in \code{amgcl::math} namespace, similar
to how backend operations are defined in \code{amgcl::backend}. Examples of
such operations are \code{math::norm()} or \code{math::adjoint()}. Arithmetic
operations like multiplication or addition are defined as operator overloads.
AMGCL algorithms at the lowest level are expressed in terms of the value type
interface, which makes it possible to transparently change the algorithm precision, or
move to complex values by adjusting the template parameter of the selected
backend.

The generic implementation of the value type operations also makes it possible
to use efficient third party implementations of the block value arithmetics.
For example, using statically sized Eigen matrices instead of the builtin
\code{amgcl::static_matrix} as block value type may improve performance in case
of relatively large blocks, since the Eigen library supports SIMD
vectorization.

\section{Stokes problem} \label{sec:stokes}

Consider the equations for the steady incompressible Stokes flow in a bounded domain $\Omega$:
\begin{equation}
    \begin{gathered}
        -\mu \nabla ^2\mathbf{u}+\nabla p = \mathbf{f}, \\
        \nabla \cdot \mathbf{u} = 0,
    \end{gathered}
    \label{eq:StokesPDE}
\end{equation}
with the following \lang{Dirichlet ($D$) and Neumann ($N$) boundary conditions}:
\begin{equation}
    \begin{gathered}
    \mathbf{u}_D=\left[ u_D,v_D \right] ,\text{ on }   \partial \Omega _D
    \\
    \mu \nabla \mathbf{u}\cdot \mathbf{n}-p\cdot \mathbf{n}=p_N\mathbf{n},\text{ on }  \partial \Omega _N
    \end{gathered}
    \label{eq:StokesBCs}
\end{equation}
where $\partial \Omega _D\cup \partial \Omega _N=\partial \Omega $ and
$\partial \Omega _D\cap \partial \Omega _N=\emptyset$; $\mu$ is the fluid dynamic
viscosity $[\si{\pascal\second}]$, $\mathbf{u}$ is the flow velocity
$[\si{\meter\per\second}]$, $p$ is the fluid pressure $[\si{\pascal}]$,
and $\mathbf{f}$ is the body force term per unit volume
$[\si{\newton\per\meter^3}]$, such as electric force and gravity.

\begin{figure}
    \centering
    \includegraphics[width=0.6\linewidth]{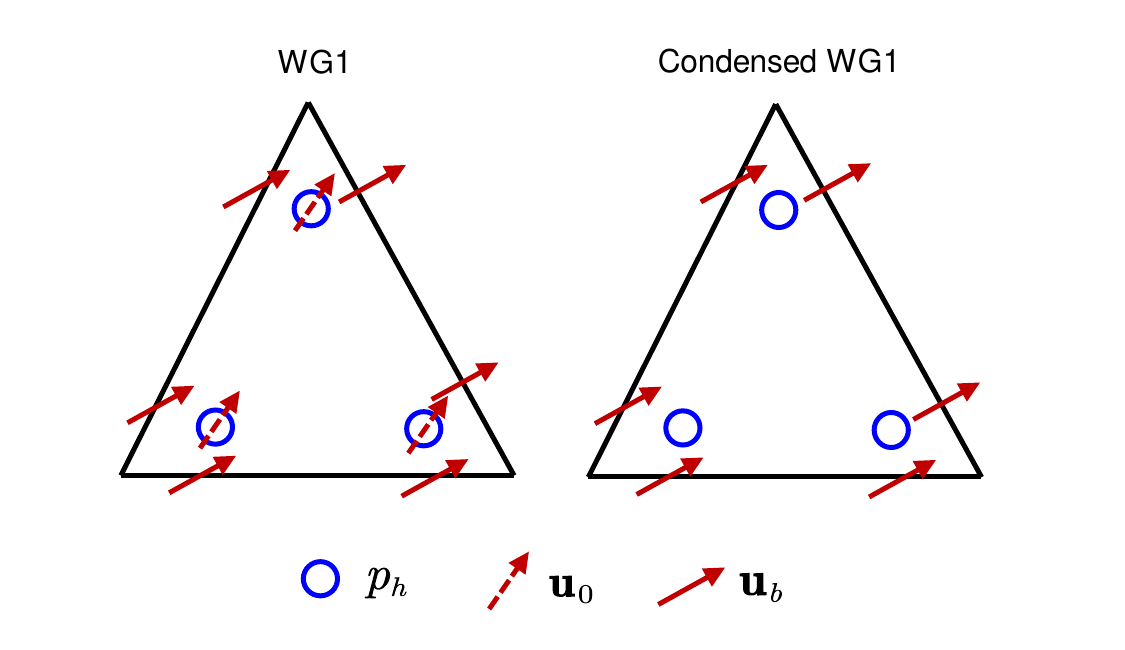}
    \caption{Illustration of the DOFs for the WGFEM method where the interior velocity DOFs can be eliminated via static condensation.}
    \label{figure:WGFEM_DOFs}
\end{figure}

The discretization of the Stokes equations~\eqref{eq:StokesPDE} is based on
a locally divergence-free weak Galerkin finite element method (WGFEM) presented
by Wang and Mu~\cite{bin2020weak}. A brief introduction of this method is given
below, see~\cite{bin2020weak} for the detailed description and analysis.
\lang{Let $T$ be a simplex element and $e=\partial T$ the element boundary.}  Denote
the finite element spaces as $V_h=\{(\mathbf{v}_0,\mathbf{v}_b):\mathbf{v}_0\in
{P}_k(T),\mathbf{v}_b\in {P}_k(e)\}$ and $W_h:=\{q\in L^2(\Omega):q\in
{P}_k(T)\}$, which are piecewise discontinuous polynomials with degree less
or equal to $k$ for any cell $T$ in the triangulation. Besides, denote the
subspace $V_h^{0,\partial\Omega_D}$ with the homogeneous boundary condition,
i.e., $\mathbf{v}_b|_{\partial\Omega_D}=0.$ A numerical approximation for
\cref{eq:StokesPDE,eq:StokesBCs} is to find
$\mathbf{u}_h=\{\mathbf{u}_0,\mathbf{u}_b\}\in V_{h}$ and $p_h\in W_h$ such
that $\mathbf{u}_b|_{\partial\Omega_D}=\mathbf{u}_D$ and
\begin{equation}
    \begin{gathered}
    \int_{\Omega}{\mu \nabla _w\mathbf{u}_h\cdot \nabla
        _w\mathbf{v}}-\int_{\Omega}{p_h\left( \nabla _w\cdot \mathbf{v}
        \right)}=\int_{\Omega}{\mathbf{f}\cdot R_T\left( \mathbf{v} \right)} \\
    +\int_{\partial \Omega _N}{p_N\mathbf{n}\cdot \mathbf{v}},\text{ }
        \forall \mathbf{v}=\left\{ \mathbf{v}_0,\mathbf{v}_b \right\} \in
        V_{h}^{0,\partial\Omega_D}, \\
    \int_{\Omega}{\left( \nabla _w\cdot \mathbf{u}_h \right) q=0},\text{ }
        \forall q\in W_h
    \end{gathered}
    \label{eq:StokesWGFEM}
\end{equation}
where  $\nabla_w$ is the weak gradient operator used in WGFEM. $R_T$ is a velocity reconstruction operator defined on the
H(div)-conforming Raviart--Thomas (RT) element space. As shown in
\Cref{figure:WGFEM_DOFs}, $p_h$ and $\mathbf{u}_0$ are the discontinuous
pressure and velocity DOFs defined on the interior of each
element. $\mathbf{u}_b$ are the continuous velocity DOFs defined on element
facets.

\begin{figure}
    \centering
    \includegraphics[width=0.7\linewidth]{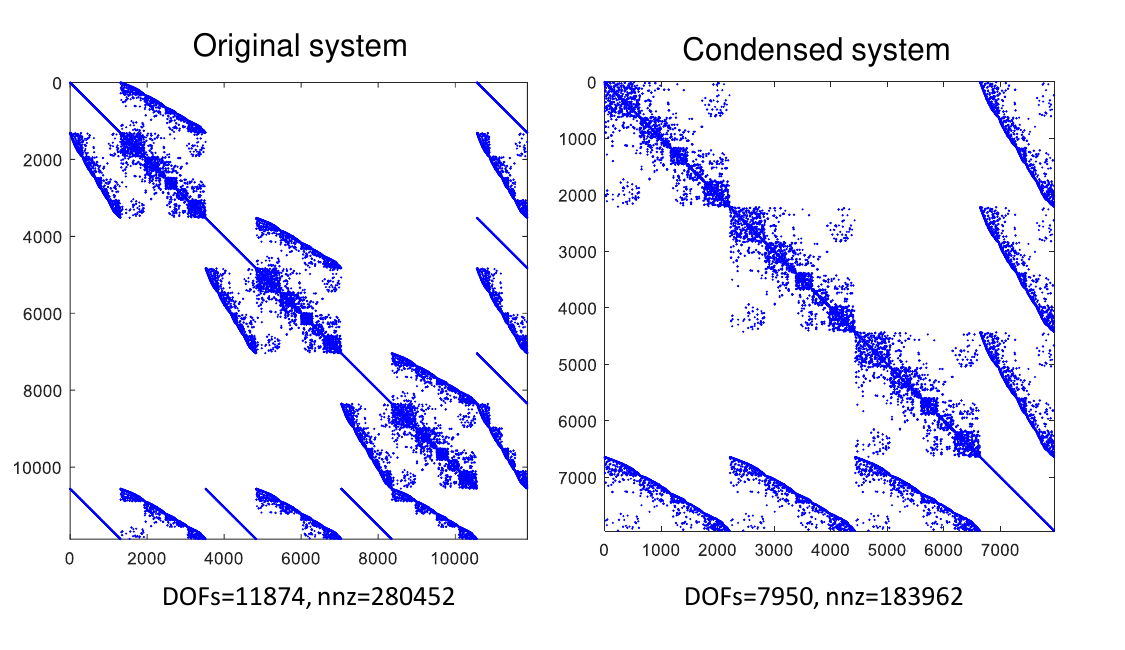}
    \caption{An example of the original and condensed linear system pattern of Stokes problem
    using WGFEM. Both linear systems are symmetric indefinite.}
    \label{figure:WGFEM_LinalgSystem}
\end{figure}

The linear system arising from \eqref{eq:StokesWGFEM} can be expressed \lang{in} block-matrix form:
\begin{equation}
    \begin{gathered}
    \left[ \begin{matrix}
	A&		B^T\\
	B&		0\\
    \end{matrix} \right] \left\{ \begin{array}{c}
    	U\\
    	P\\
    \end{array} \right\} =\left\{ \begin{array}{c}
    	L_U\\
    	L_P\\
    \end{array} \right\}
    \end{gathered}
    \label{eq:FullLinearSys}
\end{equation}
where velocity DOFs have a form of $U=\left[ U_0,U_b \right] ^T$ which
corresponds to the interior and facet velocity coefficients of $\mathbf{u}_0$
and $\mathbf{u}_b$, respectively. Since $\mathbf{u}_0$ is defined on a discontinuous
space, $A$ has a block-diagonal structure. $U_0$ can be eliminated from the full
linear system \eqref{eq:FullLinearSys} via static condensation to produce a
significantly smaller condensed system. The linear system \eqref{eq:FullLinearSys} can
be rearranged as:

\begin{equation}
    \begin{gathered}
   \left[ \begin{matrix}
	A_{00}&		A_{01}\\
	A_{01}&		A_{11}\\
    \end{matrix} \right] \left\{ \begin{array}{c}
    	X_0\\
    	X_1\\
    \end{array} \right\} =\left\{ \begin{array}{c}
    	F_0\\
    	F_1\\
    \end{array} \right\}
    \end{gathered}
    \label{eq:FullLinearSys_SC}
\end{equation}
where $X_0=\left[ U_0 \right] ^T$ and $X_1=\left[ U_b,P \right] ^T$.
Substituting $X_0=A_{00}^{-1}\left( F_0-A_{01}X_1 \right)$ into
\eqref{eq:FullLinearSys_SC} yields the condensed system:
\begin{equation}
    \left[ A_{11}-A_{10}\left( A_{00}^{-1} \right) A_{01} \right] X_1=\left\{ F_1-A_{10}\left( A_{00}^{-1} \right) F_0 \right\}.
\label{eq:Chap2_LinearSystem_SC_Final}
\end{equation}
The condensed linear system can be written in a block-matrix form as:
\begin{equation}
    \left[ \begin{matrix}
	A_c&		B_{c}^{T}\\
	B_c&		C\\
    \end{matrix} \right] \left\{ \begin{array}{c}
    	U_b\\
    	P\\
    \end{array} \right\} =\left\{ \begin{array}{c}
    	L_{Uc}\\
    	L_{Pc}\\
    \end{array} \right\}.
    \label{eq:CondensedLinearSys}
\end{equation}
The condensed linear system \eqref{eq:CondensedLinearSys} is significantly
smaller than the original system \eqref{eq:FullLinearSys}. Taking a 3D unit cube case
(\Cref{figure:WGFEM_LinalgSystem}) with a linear finite element approximation
($k=1$) as an example, the number of global DOFs is reduced by about 30\% via static
condensation. All the linear systems benchmarked in \Cref{sec:performance} are assembled from
the condensed system \eqref{eq:CondensedLinearSys} with the linear pressure and
velocity approximation ($k=1$). The system is a saddle point problem.  There
is a lot of research dedicated to the effective solution of such systems,
see~\cite{benzi2005numerical} for an extensive overview. Common approaches are
direct solvers and preconditioned Krylov subspace iterative solvers. The
following section provides an overview of the solvers studied in this work.

\section{Linear solvers for the Stokes problem} \label{sec:solvers}

\begin{lstlisting}[float=t,
    caption={IDR($s=5$) preconditioned with ILU($k=1$) (AMGCL V1)}, label=lst:V1]
typedef amgcl::backend::builtin<double> Backend;
typedef amgcl::make_solver<
  amgcl::relaxation::as_preconditioner<
    Backend,
    amgcl::relaxation::iluk
    >,
  amgcl::solver::idrs<Backend>
  > Solver;

Solver::params prm;
prm.solver.s = 5;
prm.precond.k = 1;
\end{lstlisting}

The PARDISO solver from MKL~\cite{schenk2001pardiso} is used as a
direct solver example. Specifically, the LU factorization solver for \lang{real and
nonsymmetric matrices} is used with the default parameters. The direct solver is
robust, but it does not scale well. Both the solution time and the memory
requirements grow non-linearly with the system matrix size.

 The performance of the preconditioned iterative solvers is tested with
AMGCL and PETSc libraries. The simplest choice for the preconditioner is a
monolithic one, which does not take the structure of the system into account.
We use the IDR($s$) solver~\cite{sonneveld2009idr} preconditioned with an
incomplete LU factorization with the first order
fill-in~\cite{saad2003iterative}. The implementation uses
\code{amgcl::relaxation::iluk} smoother from AMGCL wrapped into
\code{amgcl::relaxation::as_preconditioner} class. The complete definition for
the solver class is shown in \Cref{lst:V1}. The solver is labeled as
AMGCL V1 in~\Cref{sec:performance}.

A preconditioner that takes the structure of the system into account should be
a better choice performance-wise. As an example of such approach, we consider
the Schur complement pressure correction preconditioner~\cite{verfurth1984combined,
gmeiner2016quantitative}. The preconditioning step consists of solving two
linear systems:
\begin{subequations} \label{eq:SchurPC}
    \begin{align}
        S P &= L_{P_c} - B_c A_c^{-1} L_{U_c}, \label{eq:SchurP} \\
        A_c U_b &= L_{U_c} - B_c^T P.          \label{eq:SchurU}
    \end{align}
\end{subequations}
Here $S$ is the Schur complement $S = C - B_c A_c^{-1} B_c^T$.  Note that
explicitly forming the Schur complement matrix is prohibitively expensive, and
the following approximation is used to create the preconditioner
for~\eqref{eq:SchurP}:
\begin{equation} \label{eq:SchurApprox}
    \hat S = C - \diag\left( B_c \diag(A_c)^{-1} B_c^T \right).
\end{equation}
There is no need to solve the equations~\eqref{eq:SchurPC} exactly. In our
experiments, we perform a single application of the corresponding preconditioner
as an approximation to $S^{-1}$ and $A_c^{-1}$. This means that the overall
preconditioner is linear, and we can use a non-flexible iterative solver with
it.  The matrix~\eqref{eq:SchurApprox} has a simple band diagonal structure, and
a diagonal preconditioner (Jacobi in case of PETSc, and
SPAI(0)~\cite{broker2002sparse} in case of AMGCL) shows reasonable performance.
The $A_c$ matrix has more interesting structure, and we found that algebraic
multigrid with incomplete LU decomposition~\cite{saad2003iterative} as a
smoother has the best performance as a preconditioner for~\eqref{eq:SchurU}.
This approach was implemented both with AMGCL and PETSc libraries.

\lang{The} PETSc implementation uses FieldSplit preconditioner of type
\code{PC_COMPOSITE_SCHUR} with \code{LOWER} approximate block factorization
type. The preconditioner for the Schur complement is generated from an
explicitly-assembled approximation $\hat S = C - B_c \diag(A_c)^{-1} B_c^T$.
This is more expensive than~\eqref{eq:SchurApprox}, but it is the best choice
available in PETSc in terms of performance for our test problems. The Jacobi and
\lang{Geometric Algebraic Multigrid (GAMG)} preconditioners are used for~\eqref{eq:SchurP} and~\eqref{eq:SchurU}
accordingly. The GAMG preconditioner is used with the default options
\lang{(smoothed aggregation coarsening with Chebyshev relaxation on each level)}.
\Cref{lst:petsc} shows the definition of the PETSc solver.

\begin{lstlisting}[float=t,
    caption={BCGSL($L=5$) preconditioned with FieldSplit (PETSc)},
    label=lst:petsc]
KSPSetType(solver, KSPBCGSL);
KSPBCGSLSetEll(solver, 5);

KSPGetPC(solver,&prec);
PCSetType(prec, PCFIELDSPLIT);
PCFieldSplitSetType(prec, PC_COMPOSITE_SCHUR);
PCFieldSplitSetSchurPre(prec, PC_FIELDSPLIT_SCHUR_PRE_SELFP, 0);
PCFieldSplitSetSchurFactType(prec, PC_FIELDSPLIT_SCHUR_FACT_LOWER);

PCFieldSplitGetSubKSP(prec, &nsplits, &subksp);
KSPSetType(subksp[0], KSPPREONLY);
KSPSetType(subksp[1], KSPPREONLY);
KSPGetPC(subksp[0], &pc_u);
KSPGetPC(subksp[1], &pc_p);
PCSetType(pc_u, PCGAMG);
PCSetType(pc_p, PCJACOBI);
\end{lstlisting}

\begin{lstlisting}[float=t,
    caption={IDR($s=5$) preconditioned with SchurPC (AMGCL V2)}, label=lst:V2]
typedef amgcl::backend::builtin<double> Backend;

typedef amgcl::make_solver<
  amgcl::amg<
    Backend,
    amgcl::coarsening::aggregation,
    amgcl::relaxation::ilut
    >,
  amgcl::solver::preonly<Backend>
  > USolver;  // Solver for $\eqref{eq:SchurU}\label{line:v2:usolver}$

typedef amgcl::make_solver<
  amgcl::relaxation::as_preconditioner<
    Backend,
    amgcl::relaxation::spai0
    >,
  amgcl::solver::preonly<Backend>
  >  PSolver;  // Solver for $\eqref{eq:SchurP}\label{line:v2:psolver}$

typedef amgcl::make_solver<
  amgcl::preconditioner::schur_pressure_correction<
    USolver,
    PSolver
    >,
  amgcl::solver::idrs<Backend>
  > Solver;

Solver::params prm;
prm.solver.s = 5;
prm.precond.adjust_p = 1; $\label{line:v2:adjust}$
\end{lstlisting}

The AMGCL implementation (labeled AMGCL V2 in~\Cref{sec:performance}) is
shown in~\Cref{lst:V2}\lang{, where} line~\ref{line:v2:adjust} enables
the approximation~\eqref{eq:SchurApprox} for the Schur complement in the
preconditioner for the system~\eqref{eq:SchurP}.  This version of the solver
has reasonable performance already, but there is still some room for
improvement.  Note that the nested solvers for systems~\eqref{eq:SchurP}
and~\eqref{eq:SchurU} in lines~\ref{line:v2:psolver} and~\ref{line:v2:usolver}
of \Cref{lst:V2} explicitly specify the backend they use. This may seem like redundancy in the library API, but in fact it enables the solver tuning.
The matrix $A_c$ has a block structure with $3 \times
3$ blocks, where each block corresponds to a single tetrahedron face.
We can exploit this with the value type system of AMGCL.
The modified nested solver declaration for~\eqref{eq:SchurU} is shown in
\Cref{lst:V3}. Namely, we declare a block-valued backend with
\code{amgcl::static_matrix<3,3,double>} as the value type, and use it for the solver. We
have to use \code{amgcl::make_block_solver<>} instead of
\code{amgcl::make_solver<>} to automatically convert the matrix $A_c$ to
the block format during setup and mix the scalar and block-valued vectors
during the preconditioning step. The resulting solver is labeled as AMGCL V3
in~\Cref{sec:performance}.
Switching to the block-valued backend has several advantages. First, the block
representation of the matrix is more efficient memory-wise. This is
demonstrated on the example of the following sparse matrix that has a $2
\times 2$ block structure:
\begin{equation*}
    \begin{bmatrix}
        0.71 & 0.65 & 0.26 & 0.79 &      &      \\
        0.54 & 0.37 & 0.17 & 0.62 &      &      \\
             &      & 0.89 & 0.05 &      &      \\
             &      & 0.27 & 0.15 &      &      \\
             &      &      &      & 0.52 & 0.34 \\
             &      &      &      & 0.45 & 0.64
    \end{bmatrix}
\end{equation*}
Below is the standard scalar representation of the matrix in CSR format:

\begin{Verbatim}[samepage=true]
ptr=[0, 4, 8, 10, 12, 14, 16]
col=[0, 1, 2, 3, 0, 1, 2, 3, 2, 3, 2, 3, 4, 5, 4, 5]
val=[0.71, 0.65, 0.26, 0.79, 0.54, 0.37, 0.17, 0.62,
     0.89, 0.05, 0.27, 0.15, 0.52, 0.34, 0.45, 0.64]
\end{Verbatim}
Compare it to the block-valued representation of the same matrix:
\begin{Verbatim}[samepage=true]
ptr=[0, 2, 3, 4]
col=[0, 1, 1, 2]
val=[[0.71, 0.65; 0.54, 0.37], [0.26, 0.79; 0.17, 0.62],
     [0.89, 0.05; 0.27, 0.15], [0.52, 0.34; 0.45, 0.64]]
\end{Verbatim}
With the block representation, the matrix has
twice fewer rows and columns \lang{(because each row and column now
spans two scalar unknowns)}, and four times fewer logical non-zero values.
This means that the matrix representation needs twice less memory to store the
\code{ptr} array, and four times less memory for the \code{col} array. The
reduced logical size of the matrix also speeds up the algorithm setup.
Finally, the block-valued backend improves cache efficiency
during the solution phase. Our experiments show that the setup phase is
around 85\% faster, the memory footprint is reduced by 15\% to 30\%, and the
overall speedup of 30\% to 40\% is achieved over the V2 solver.

\begin{lstlisting}[float=t,
    caption={Nested solver declaration for~\eqref{eq:SchurU} using
    the block-valued backend (AMGCL V3)},
    label=lst:V3]
typedef amgcl::backend::builtin<amgcl::static_matrix<3,3,double>> BlockBackend;
typedef amgcl::make_block_solver<
  amgcl::amg<
    BlockBackend,
    amgcl::coarsening::aggregation,
    amgcl::relaxation::ilut
    >,
  amgcl::solver::preonly<BlockBackend>
  > USolver;
\end{lstlisting}

\begin{lstlisting}[float=t,
    caption={Declaration of single-precision backends for the nested solvers in
    \Cref{lst:V2} (AMGCL V4)},
    label=lst:V4]
typedef amgcl::backend::builtin<amgcl::static_matrix<3,3,float>> BlockBackend;
typedef amgcl::backend::builtin<float> SBackend;
\end{lstlisting}

Another possibility to improve the solver performance is to use the mixed
precision approach. A preconditioner only needs to approximate
the inverse of the system matrix, which makes the single precision values
a natural choice for the preconditioning step. It is enough
to use the single precision valued backends with the nested solvers
for~\eqref{eq:SchurPC}, as shown in~\Cref{lst:V4}.  The outer iterative
solver still uses the double precision backend which makes sure the final
result has the desired accuracy.  The mixed precision version of the solver
is labeled as AMGCL V4 in~\Cref{sec:performance}. As shown in our
experiments, using the mixed-precision approach does not increase the number of
iterations, \lang{and} further reduces the memory footprint
of the algorithm by around 30\% and speeds up the solution phase by 30\% to
40\% with respect to the V3 solver. The latter is explained by the fact that
most iterative solvers are memory-bound, and switching to the
single precision effectively doubles the available memory bandwidth.

Note that AMGCL V2, V3, and V4 solvers use the same source
code on all stages of the solution, and only differ by the value type of the
nested solver backends. The \CXX templating system combined with the value type
concept in AMGCL allows the compiler to generate efficient machine code for each
of the cases.

\section{Performance study} \label{sec:performance}

In this section, we evaluate the performance of the direct MKL Pardiso
solver v2020 and the five iterative solvers introduced in \Cref{sec:stokes} and
listed in \Cref{table:IterSolvers}. The solvers are benchmarked using the three
problems described in the following subsections. All tests \lang{were} conducted on a
workstation with 3.30GHz 10-core Intel I9-9820X processor and 64GB of RAM. The
Pardiso and the AMGCL solvers \lang{use} thread parallelism with the OpenMP
technology, and the PETSc solver is parallelized using MPI with a single MPI
process per CPU core. The source code for the benchmarks is published
\lang{in}~\cite{cppstokesbm}, and the dataset used in the benchmarks is 
available \lang{in}~\cite{cppstokesds}.  The AMGCL tests were compiled with
\code{g++} v9.2.0 compiler with the \code{O3} optimization level. The PETSc
tests were linked against PETSc v3.14.0.

\begin{table}
\caption{Iterative solvers used in the performance tests.}
\centering
\begin{tabular}{clllc}
\hline
         & Solver   & Preconditioner & Value type  & Mixed Precision \\ \hline
AMGCL V1 & IDR(5)   & ILU(1)         & Scalar      & No              \\
AMGCL V2 & IDR(5)   & SchurPC        & Scalar      & No              \\
AMGCL V3 & IDR(5)   & SchurPC        & Block (3x3) & No              \\
AMGCL V4 & IDR(5)   & SchurPC        & Block (3x3) & Yes             \\
PETSc    & BCGSL(5) & FieldSplit     & Scalar      & No              \\
\hline
\end{tabular}
\label{table:IterSolvers}
\end{table}

\subsection{Unit cube problem} \label{sec:test1}

\begin{figure}
    \centering
    \includegraphics[width=0.44\linewidth]{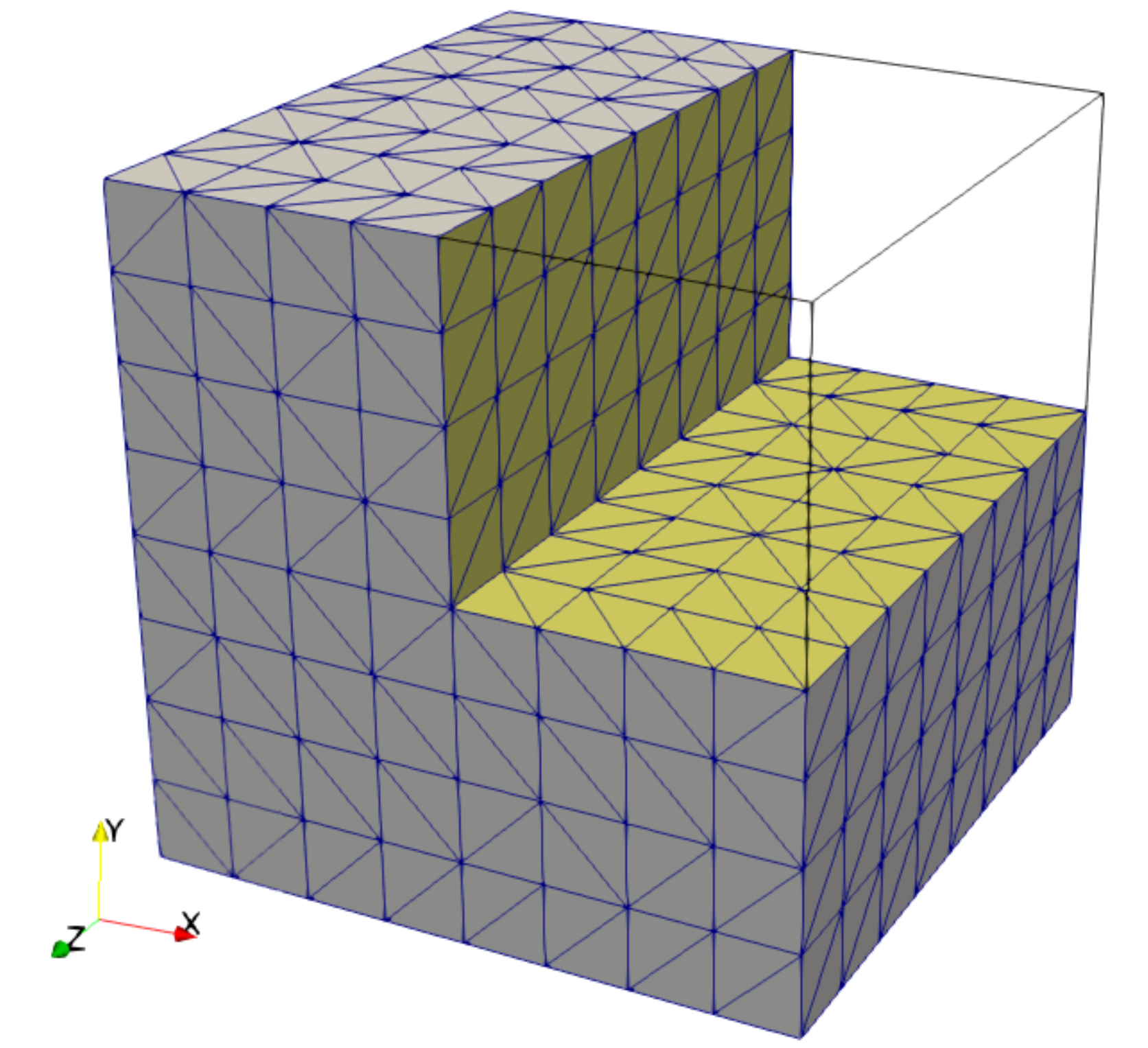}
    \includegraphics[width=0.49\linewidth]{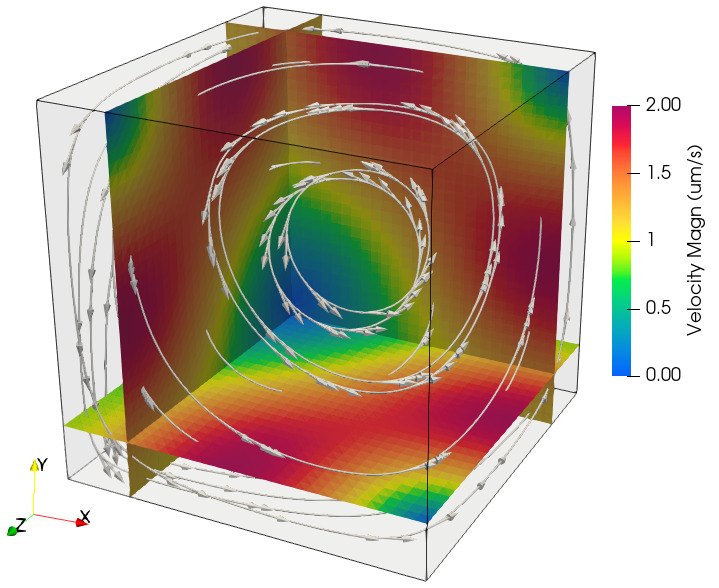}
    \caption{Mesh and velocity field for the unit cube problem}
    \label{figure:Case1_mesh}
\end{figure}

Consider a rotating flow driven by an external force $\mathbf{f}$ in a closed unit cube
$\Omega=[0,1]^3$ with the constant viscosity $\mu=1$ as in
\cite{rhebergen2020embedded}. The external force $\mathbf{f}$ and the analytical
solution $\mathbf{u}$ and $p$ for this problem can be expressed as:
\begin{equation}
    \begin{gathered}
            \mathbf{f}=\pi \left[ \begin{aligned}
                 2\pi \sin \left( \pi x \right) \cos \left( \pi y \right) &- 2\pi \sin \left( \pi x \right) \cos \left( \pi z \right) + \sin \left( \pi y \right) \sin \left( \pi z \right) \cos \left( \pi x \right)\\
                -2\pi \sin \left( \pi y \right) \cos \left( \pi x \right) &+ 2\pi \sin \left( \pi y \right) \cos \left( \pi z \right) + \sin \left( \pi x \right) \sin \left( \pi z \right) \cos \left( \pi y \right)\\
                 2\pi \sin \left( \pi z \right) \cos \left( \pi x \right) &- 2\pi \sin \left( \pi z \right) \cos \left( \pi y \right) + \sin \left( \pi x \right) \sin \left( \pi y \right) \cos \left( \pi z \right)\\
            \end{aligned} \right],
            \\
           \mathbf{u}=\pi \left[ \begin{array}{c}
                \sin \left( \pi x \right) \cos \left( \pi y \right) -\sin \left( \pi x \right) \cos \left( \pi z \right)\\
                \sin \left( \pi y \right) \cos \left( \pi z \right) -\sin \left( \pi y \right) \cos \left( \pi x \right)\\
                \sin \left( \pi z \right) \cos \left( \pi x \right) -\sin \left( \pi z \right) \cos \left( \pi y \right)\\
            \end{array} \right],
            \\
            p=\sin \left( \pi x \right) \sin \left( \pi y \right) \sin \left( \pi z \right) -\frac{8}{\pi ^3}.
    \end{gathered}
\label{eq:UnitCubeProb}
\end{equation}
The domain is partitioned into an unstructured tetrahedral mesh in 5 levels of
refinement using Gmsh~\cite{geuzaine2009gmsh}. \lang{The mesh and the computed velocity field
are shown in~\Cref{figure:Case1_mesh}.} The number of elements,
DOFs, non-zeros, and the memory occupied by the system matrix in the CSR format on
each level is presented in \Cref{table:UnitCube}.

\begin{table}
\caption{Linear system size (DOFs), number of non-zeros (NNZs), and
    linear system CSR storage memory for the unit cube problem.}
\centering
\begin{tabular}{rrrrrr}
\hline
Cells   & DOFs      & NNZs        & CSR Memory \\ \hline
    384 &     9~312 &     195~000 &   6.40 Mb  \\
  3~072 &    71~040 &   1~710~000 &  52.64 Mb  \\
 24~576 &   554~496 &  14~300~000 & 427.65 Mb  \\
131~712 & 2~940~000 &  78~400~000 &   2.27 Gb  \\
196~608 & 4~380~672 & 117~429~672 &   3.39 Gb  \\
\hline
\end{tabular}
\label{table:UnitCube}
\end{table}

\begin{figure}
    \centering
    \includegraphics[width=\linewidth]{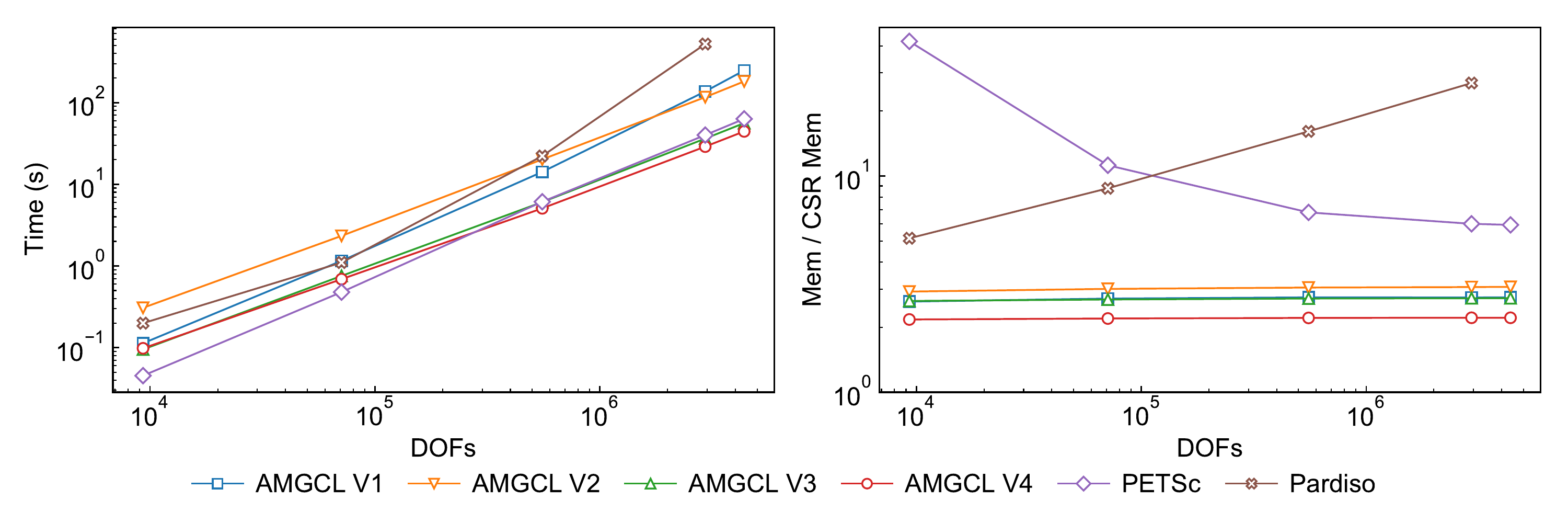}
    \includegraphics[width=0.48\linewidth]{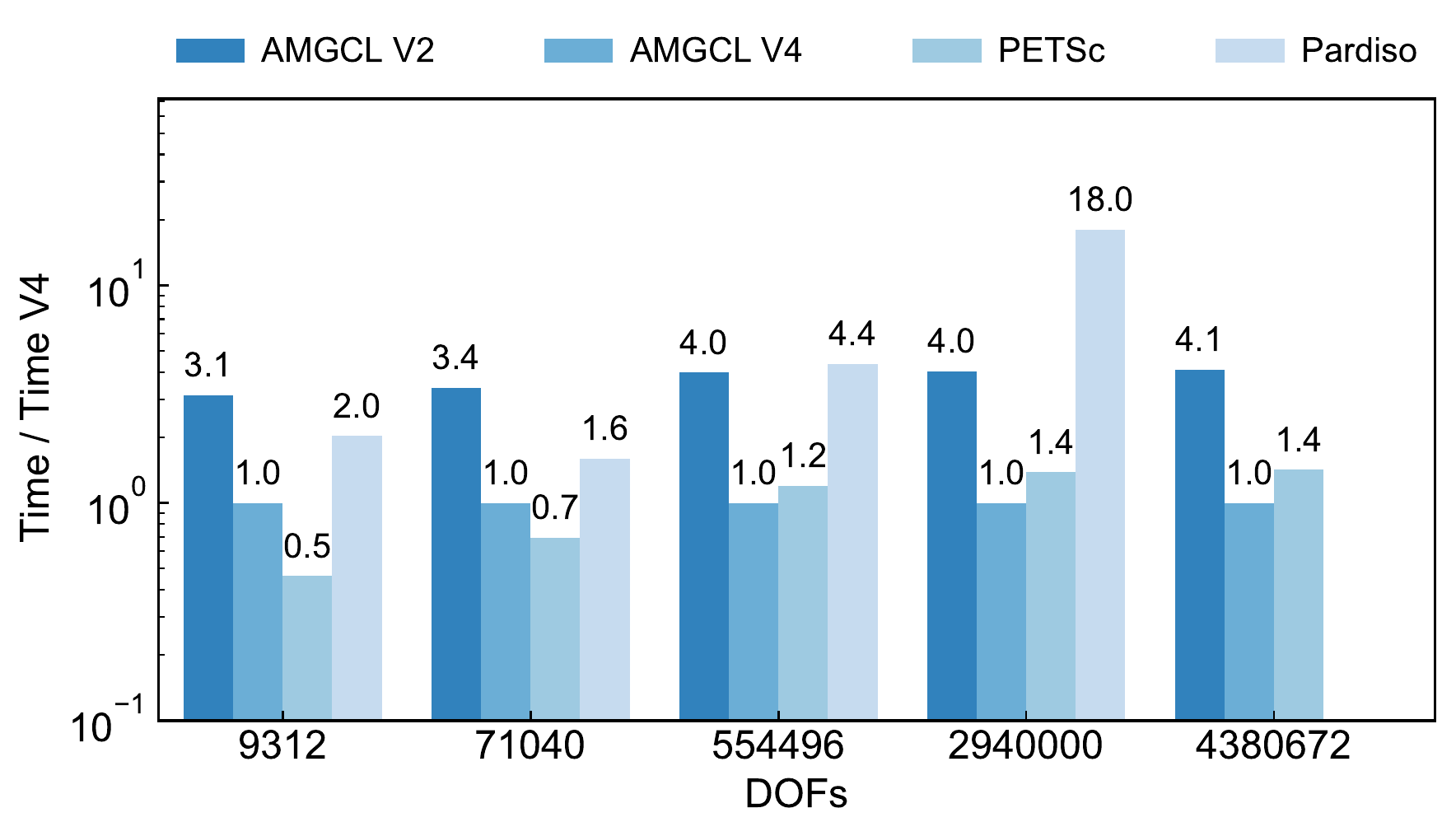}
    \includegraphics[width=0.48\linewidth]{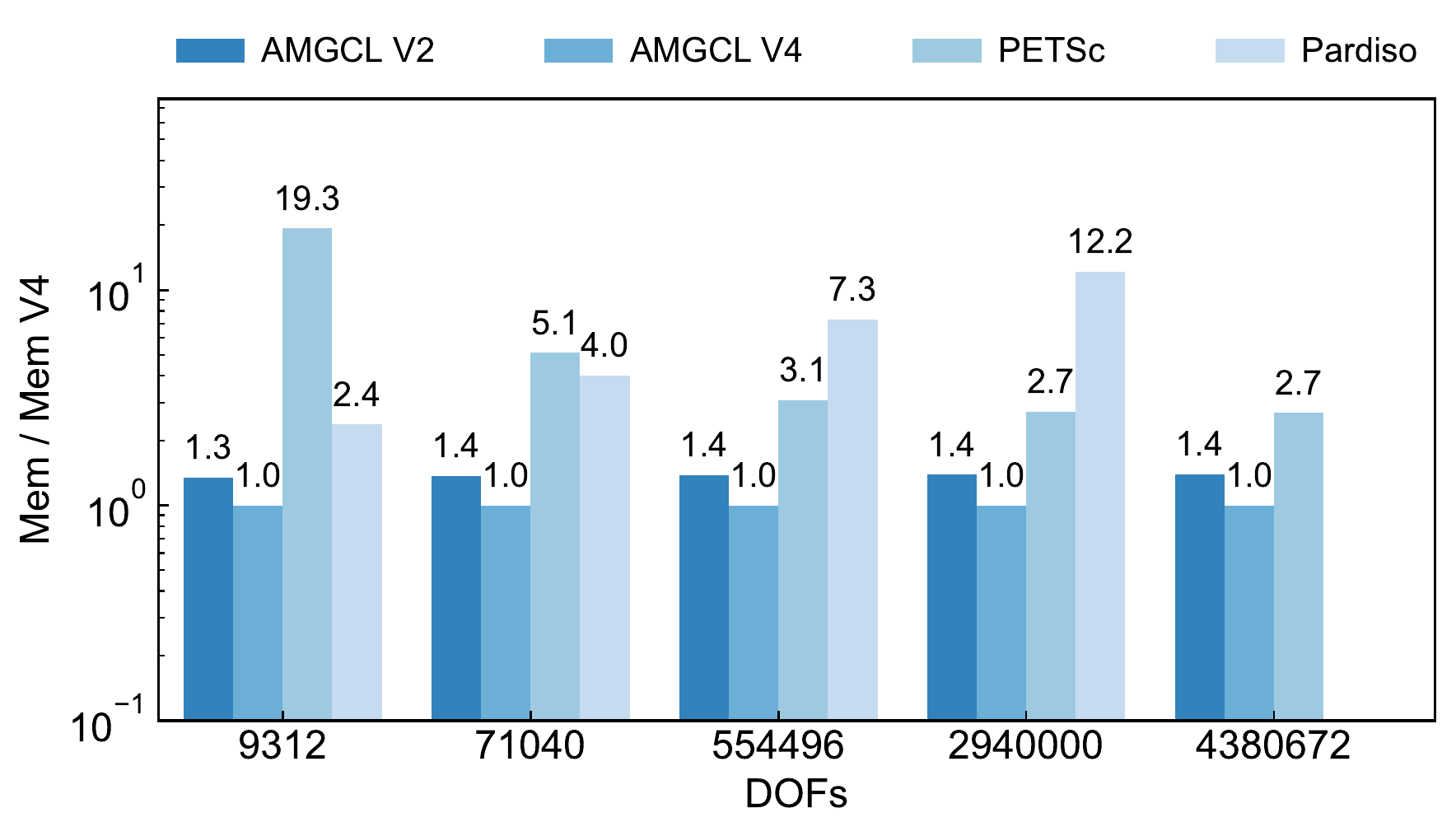}
    \caption{Solution time and memory usage for the unit cube problem.}
    \label{figure:Case1_Performance}
\end{figure}

\Cref{figure:Case1_Performance} shows the solution time and the memory
usage for the AMGCL, PETSc, and Pardiso solvers. The top left plot shows the
solution time in seconds, the top right plot shows the memory footprint of each
solver scaled by the size of the system CSR matrix. The plots in the bottom row
present the solution time and the memory footprint scaled by the AMGCL V4
solver results. Here the results for AMGCL V1 and V3 are omitted for the sake
of readability.

As expected, the direct Pardiso solver scales worse than any of the
iterative solvers with respect to both the solution time and especially the
memory footprint. In fact, the memory requirements of the Pardiso are so high
that it was not able to process the largest problem ($4.4\times 10^6$ DOFs) on
the workstation with 64~Gb of RAM. The system with $2.9\times10^6$~DOFs
consumed 61.1~Gb with the Pardiso solver, as opposed to 5.1~Gb with the AMGCL
V4 solver.

Among the iterative solvers, the monolithic AMGCL V1 appears to be the
least scalable. The convergence rate of AMGCL V1 deteriorates with the problem
size and the number of iterations increases from 20 for the smallest problem to
1004 for the largest one. However, the memory requirements of this simple
method are low (comparable to AMGCL V3), and each iteration is cheap with
respect to the composite preconditioners, which allows AMGCL V1 to
outperform V2 for problems smaller than $2.9\times10^6$~DOFs.

All solvers using the Schur pressure correction preconditioner show
similar scalability in terms of the compute time. The PETSc version has the
best performance for problems below $0.5\times10^6$~DOFs, and consistently
outperforms AMGCL V2 for all problem sizes. Memory-wise, the PETSc version
apparently has a constant overhead of around 200~Mb, which explains its
relatively high memory usage for the smaller problems. AMGCL solvers stay the
most memory-efficient even for the largest problem, where PETSc uses 20.6~Gb,
and AMGCL solvers consume from 7.6~Gb (V4) to 10.6~Gb (V2). AMGCL V3 and V4
outperform PETSc for the three largest problems. The switch to the
block-valued backend for~\eqref{eq:SchurU} has the largest impact on
performance, since the V3 and V4 solution times are very close. AMGCL V4
is up to 4.1 times faster than V2 and only about 25\% faster than V3. Due
to the mixed precision approach, V4 memory requirements are 40\% less than V2,
and 25\% less than V1 and V3. Although the AMGCL V2, V3, and V4 solvers use exactly
the same code, the block-valued backend combined with the mixed precision
approach allows the AMGCL V4 to outperform PETSc by about 40\% for the
largest problem sizes.

\subsection{Converging-diverging tube problem} \label{sec:test2}

\begin{figure}
    \centering
    \includegraphics[width=0.45\linewidth]{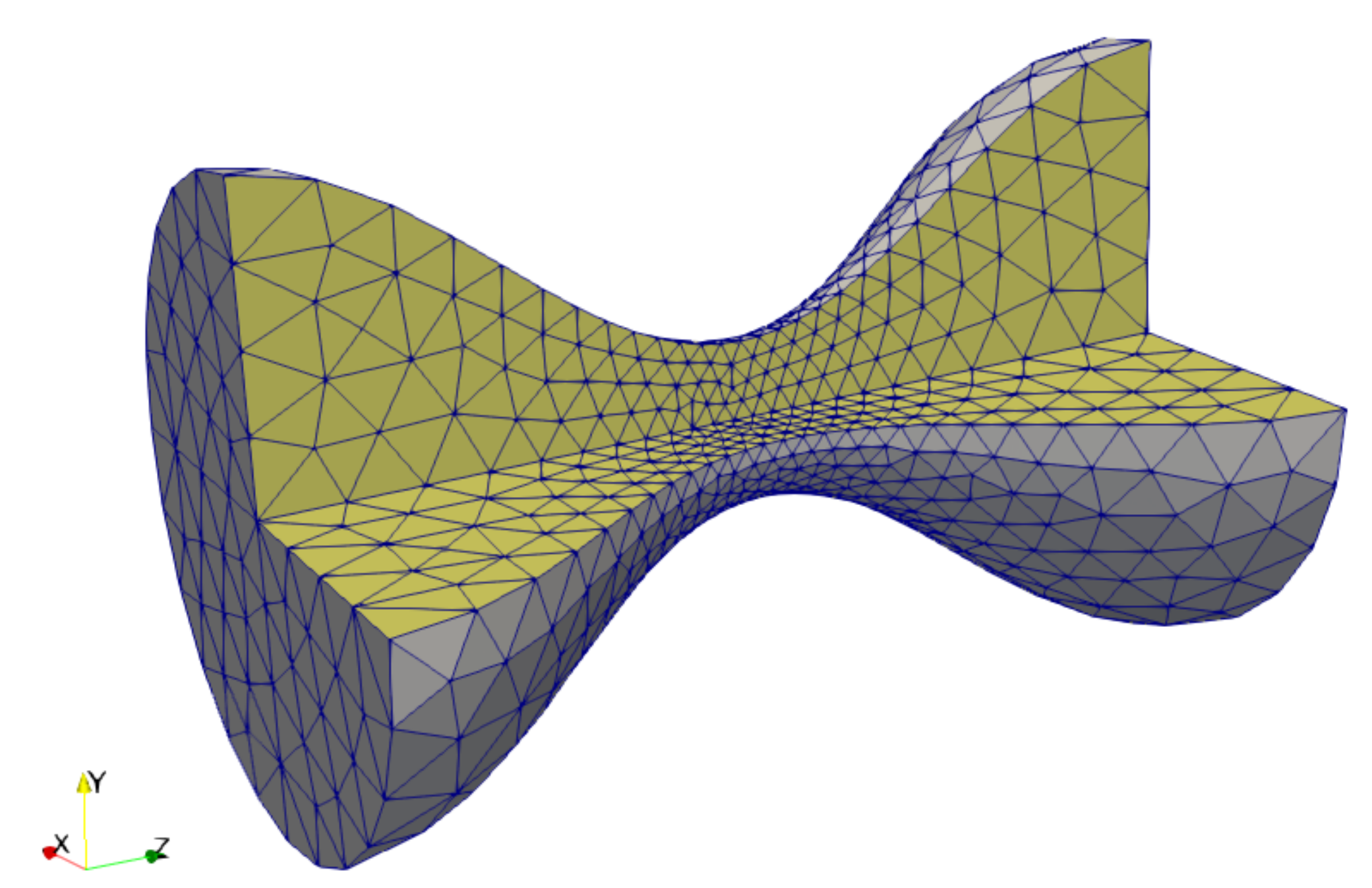}
    \includegraphics[width=0.45\linewidth]{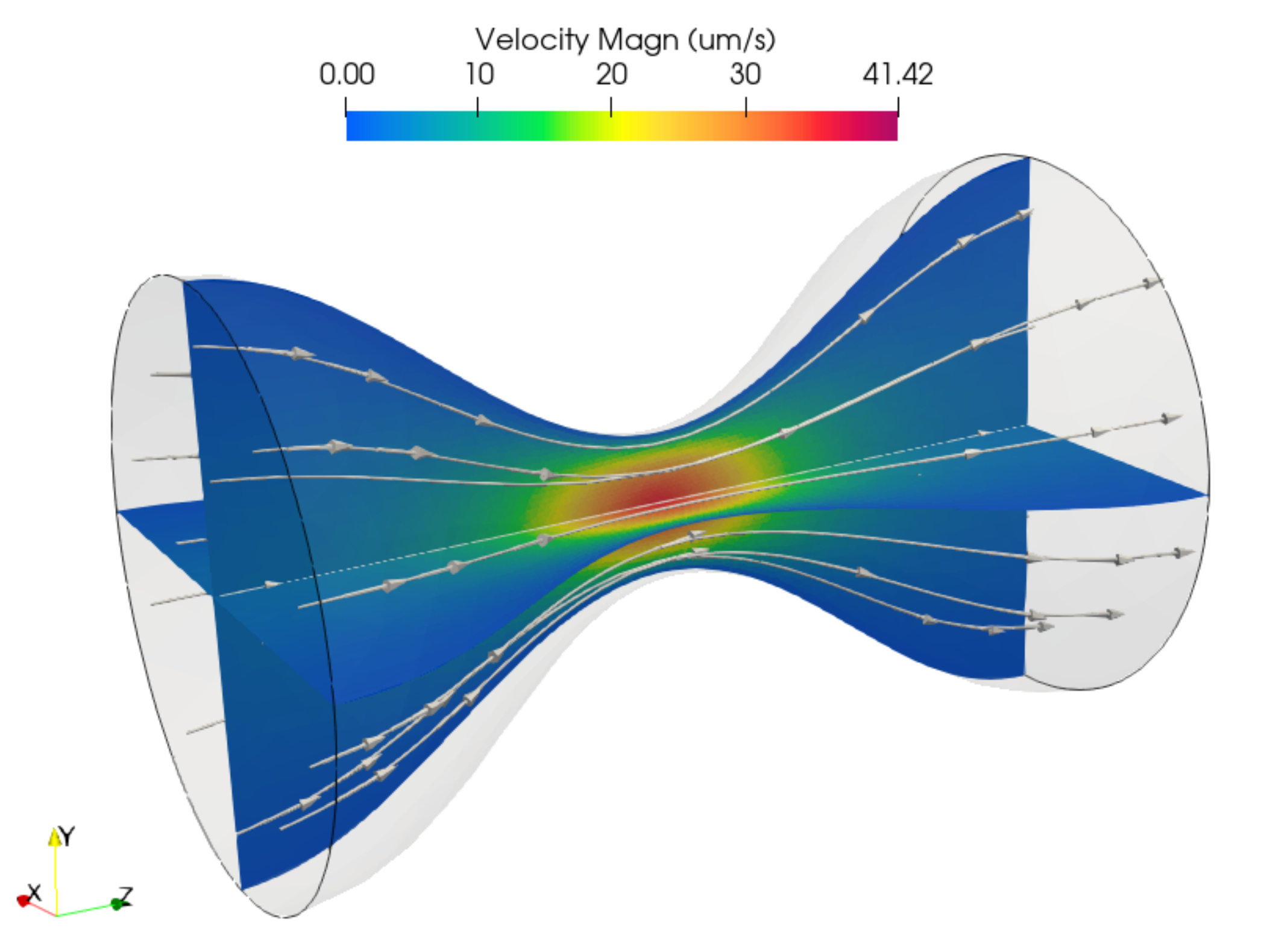}
    \caption{Mesh and velocity field for the converging-diverging tube problem}
    \label{figure:Case2_mesh}
\end{figure}

\begin{figure}
    \centering
    \includegraphics[width=\linewidth]{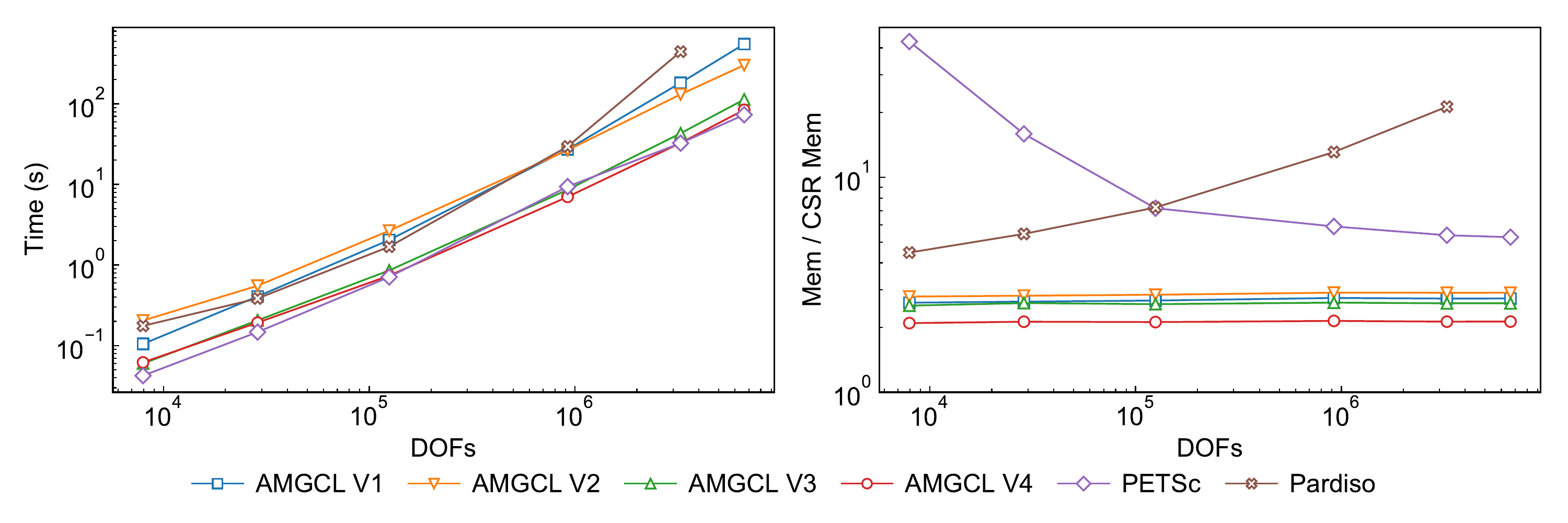}
    \includegraphics[width=0.48\linewidth]{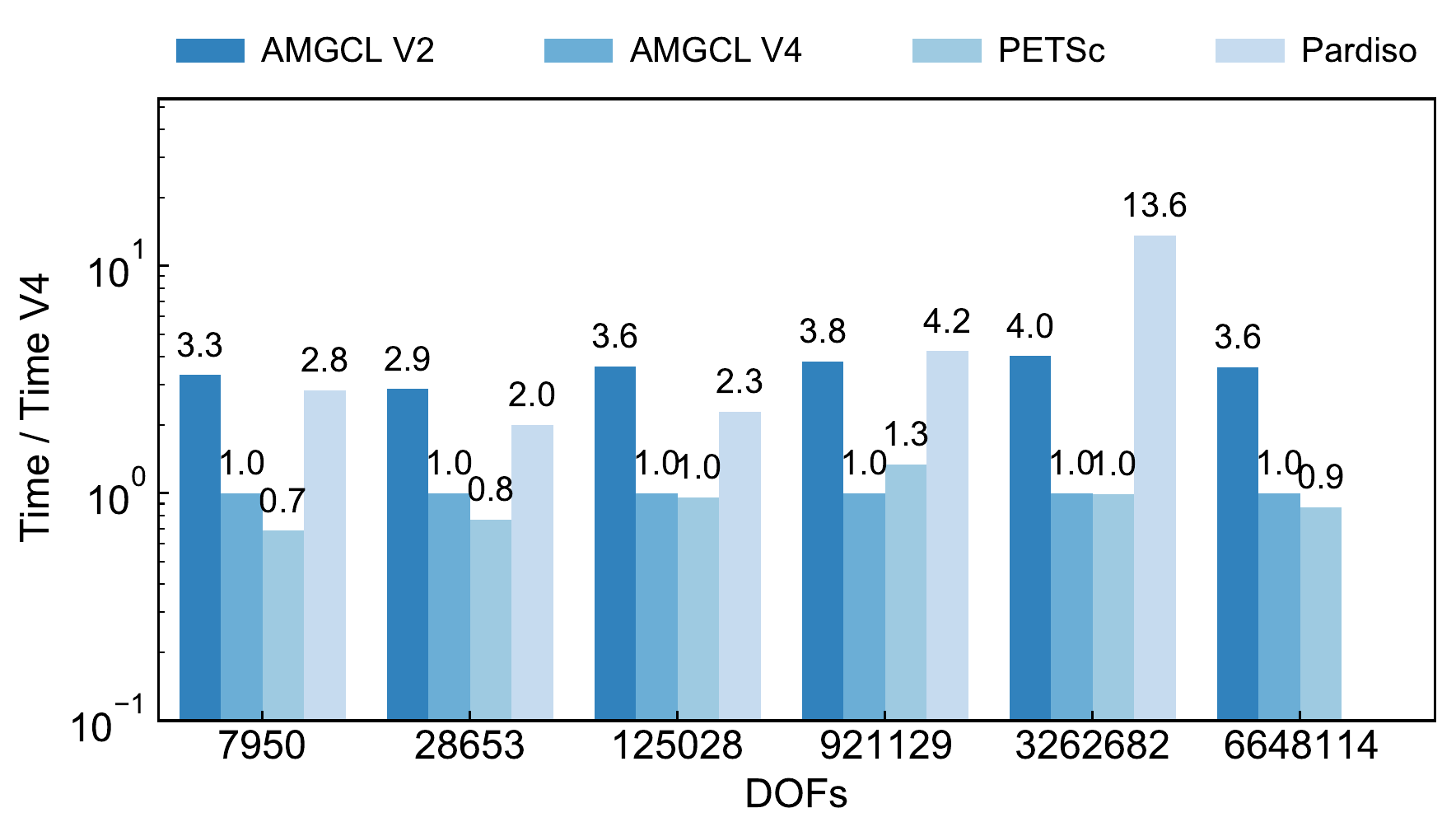}
    \includegraphics[width=0.48\linewidth]{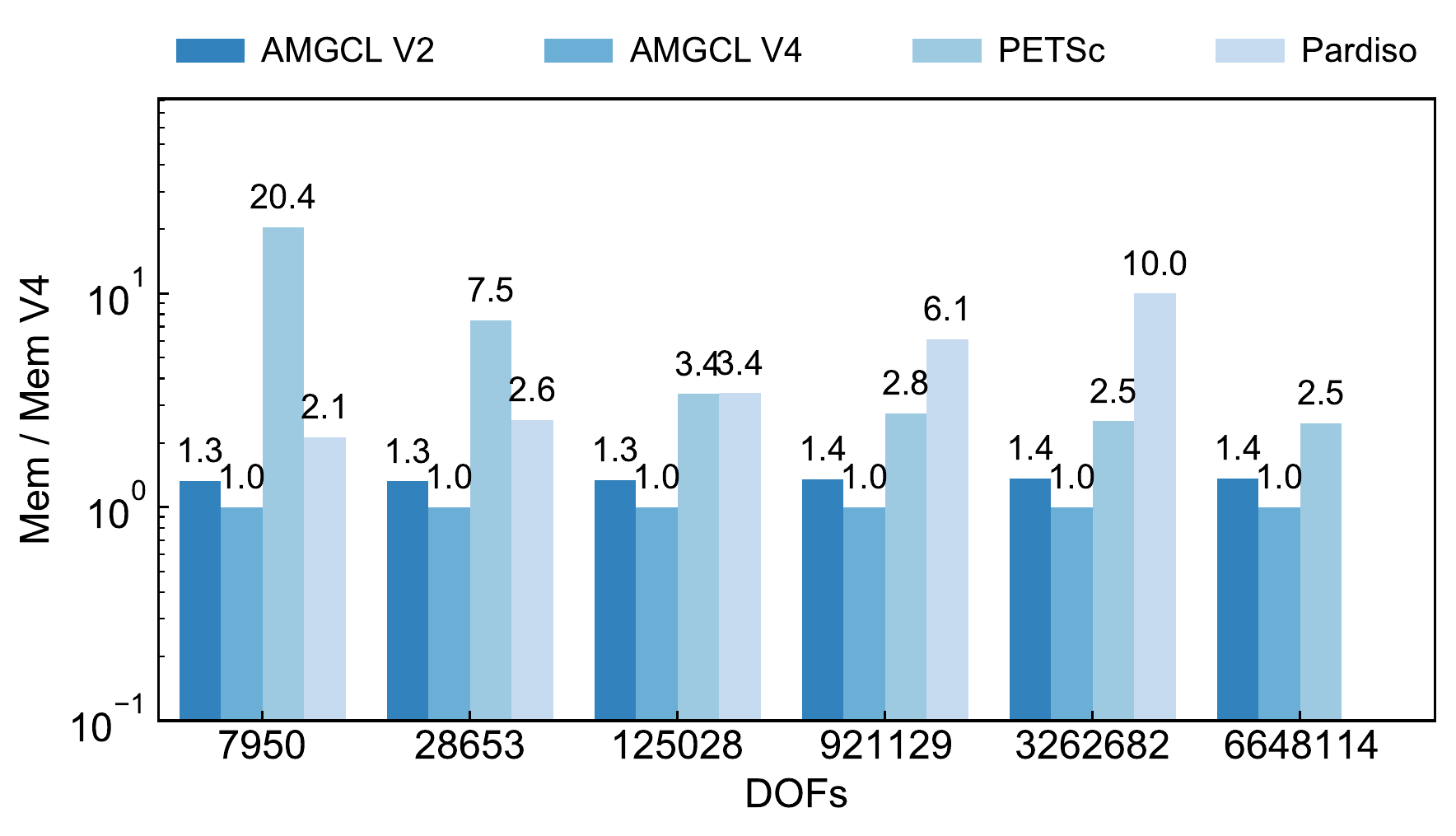}
    \caption{Solution time and memory usage for the converging-diverging
    tube problem.}
    \label{figure:Case2_Performance}
\end{figure}

In this case, the solution performance is evaluated with the example of
a pressure-driven tube flow (\Cref{figure:Case2_mesh}). Consider the Stokes flow
through a 3D converging-diverging tube under a pressure drop of
$1\si{\pascal}$. The Neumann boundary conditions are imposed at the top ($z =
-3.75\si{\micro\metre}$) and the bottom ($z = 1.25\si{\micro\metre}$) walls. The
no-slip wall boundary conditions of \lang{$\mathbf{u}=\left(0,0,0\right)^T$} are
imposed on the surface of the varying radius tube. The fluid dynamics viscosity is
uniformly set as $1\si{\pascal\second}$.  Given the tube length
$L=5\si{\micro\metre}$ and the average tube radius of $R_0=1\si{\micro\metre}$, the
radius of the axisymmetric tube can be expressed as:
\begin{equation}
    R\left( z \right) =1.0+0.6 \sin \left( \frac{2\pi z}{5.0} \right).
    \label{eq:Case2_TubeGeometry}
\end{equation}
The 6 mesh refinement levels are generated using Gmsh. \lang{The mesh and the
computed velocity field are shown in~\Cref{figure:Case2_mesh}.}
\Cref{table:ConDivTube} shows the problem size at each level\lang{, and}
\Cref{figure:Case2_Performance} presents the solution time and the memory usage
scaled by the CSR matrix size and by the AMGCL V4 solver results.

\begin{table}
\caption{Linear system size (DOFs), number of non-zeros (NNZs), and
    linear system CSR storage memory for the converging-diverging tube problem.}
\centering
\begin{tabular}{rrrrrr}
\hline
Cells   & DOFs      & NNZs        & CSR Memory \\ \hline
327     & 7~950     & 184~000     & 6.07 Mb    \\
1~203   & 28~653    & 710~000     & 22.78 Mb   \\
5~319   & 125~028   & 3~220~000   & 101.43 Mb  \\
40~488  & 921~129   & 25~300~000  & 766.66 Mb  \\
144~995 & 3~262~682 & 95~300~000  & 2.81 Gb    \\
296~823 & 6~648~114 & 196~657~124 & 5.78 Gb    \\
\hline
\end{tabular}
\label{table:ConDivTube}
\end{table}

The main trends here are similar to the unit cube problem case. The direct
Pardiso solver shows the worst scaling both in solution time and memory, and is
not able to solve the largest system of $6.6\times10^6$~DOFs due to the memory
limitations. The iterative AMGCL V1 solver with a simple monolithic
preconditioner has low memory usage but scales badly
with respect to the solution time. Both Pardiso and the AMGCL V1 solvers are
outperformed by AMGCL V2 starting with the $0.9\times10^6$~DOFs problem.

The AMGCL V3 and V4 solvers, and the PETSc solver show similar performance,
although this time PETSc has slightly lower solution time everywhere except
the $0.9\times10^6$~DOFs problem, where it is outperformed by both AMGCL V3 and
V4. Again, PETSc has a large constant memory overhead of approximately
250~Mb, and is overall more expensive memory-wise. At the largest problem of
$6.6\times10^6$~DOFs PETSc needs 31.2~Gb, while AMGCL only requires
from 12.6~Gb (V4) to 17.2~Gb (V2). AMGCL V4 is 3--4 times faster than V2 and is
consistently the most memory-efficient solver.

\subsection{Sphere packing problem} \label{sec:test3}

\begin{figure}
    \centering
    \includegraphics[width=0.38\linewidth]{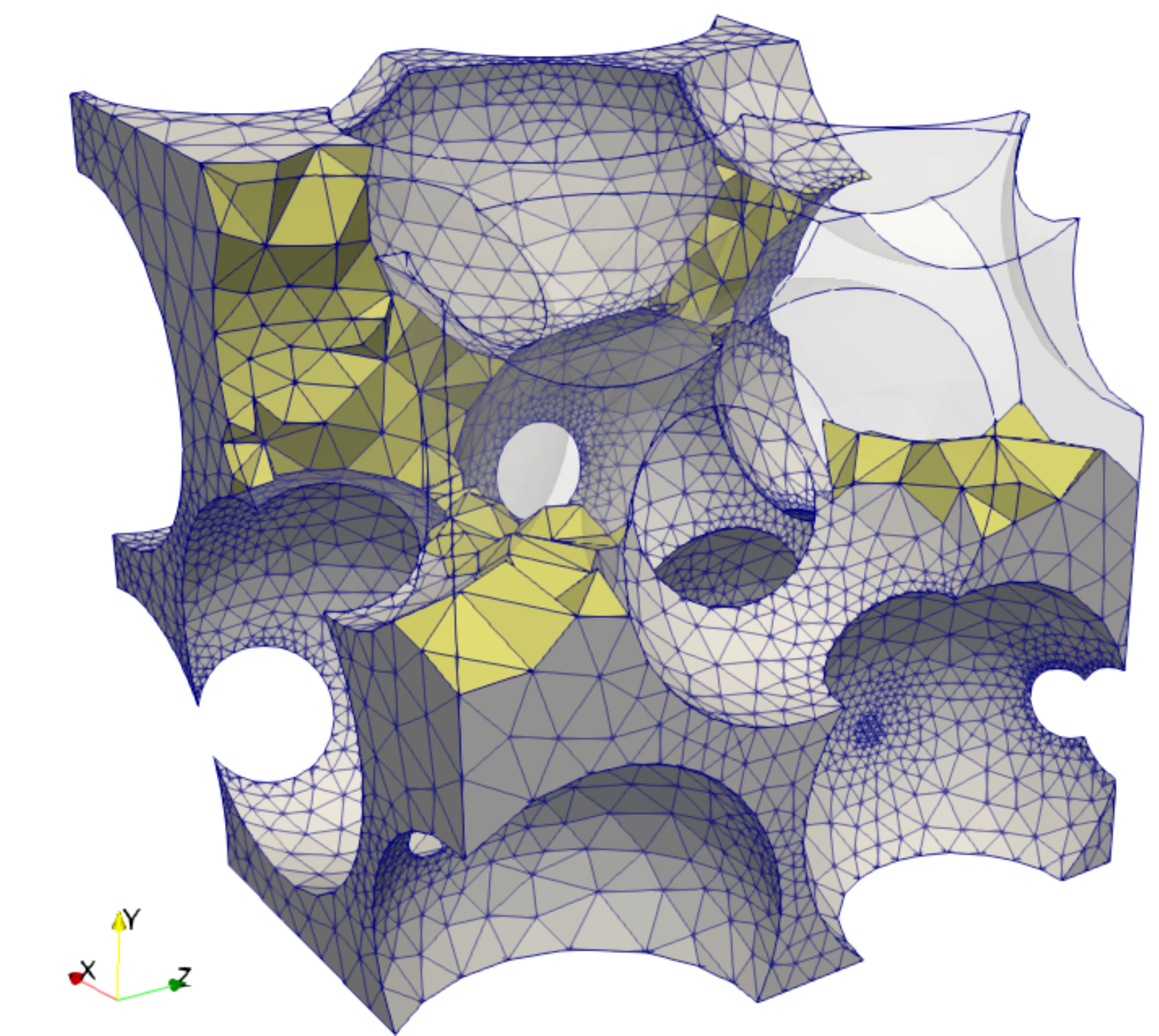}
    \includegraphics[width=0.44\linewidth]{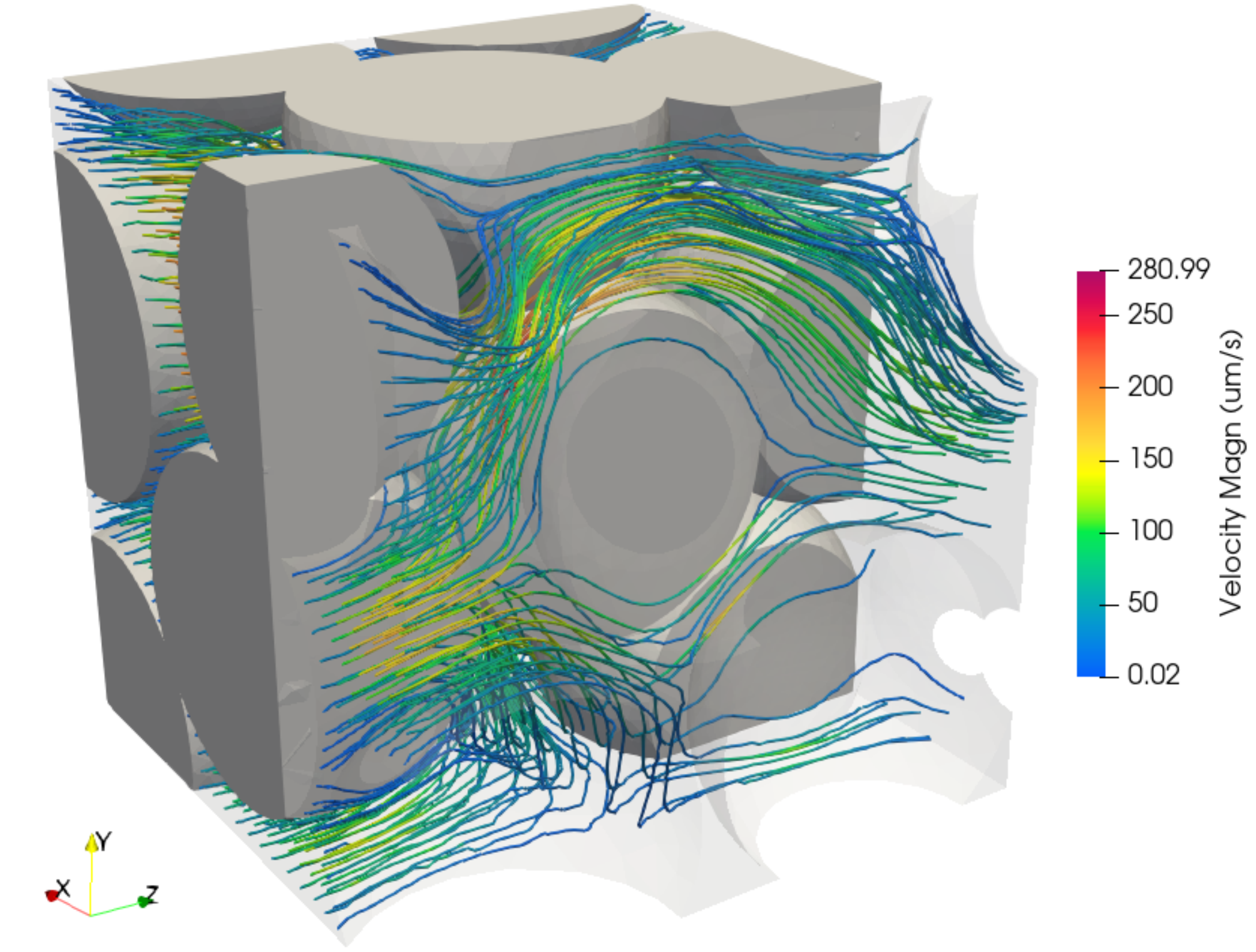}
    \caption{Mesh and velocity field for the sphere packing problem}
    \label{figure:Case3_mesh}
\end{figure}

\begin{figure}
    \centering
    \includegraphics[width=\linewidth]{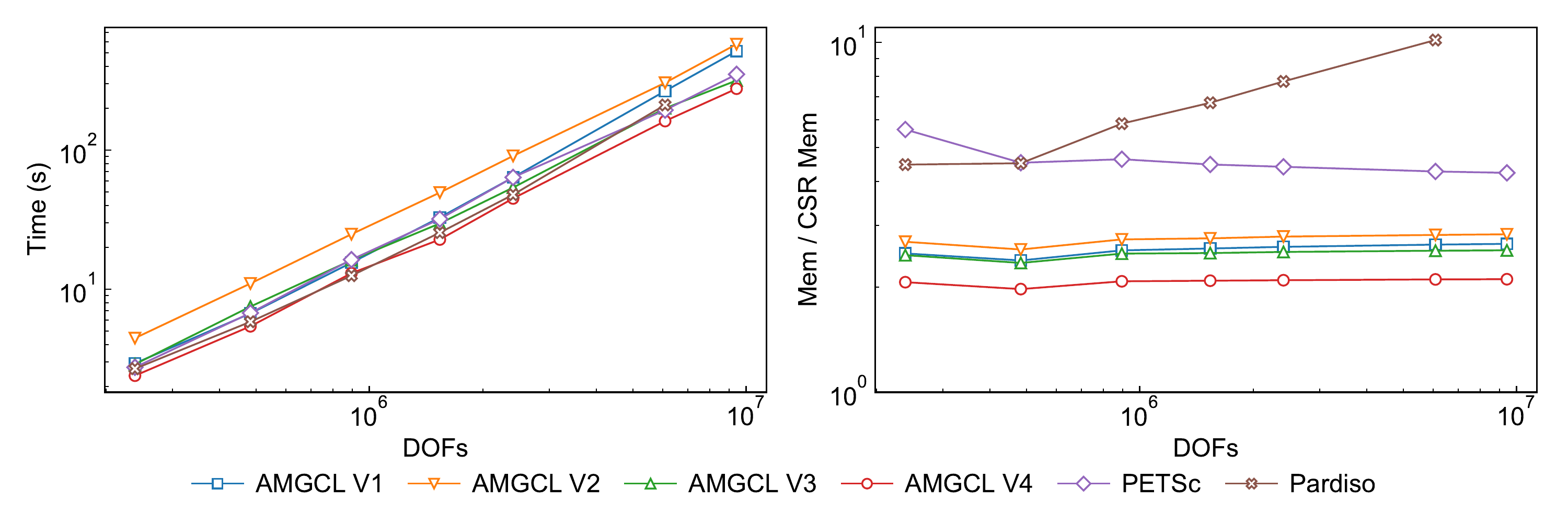}
    \includegraphics[width=0.48\linewidth]{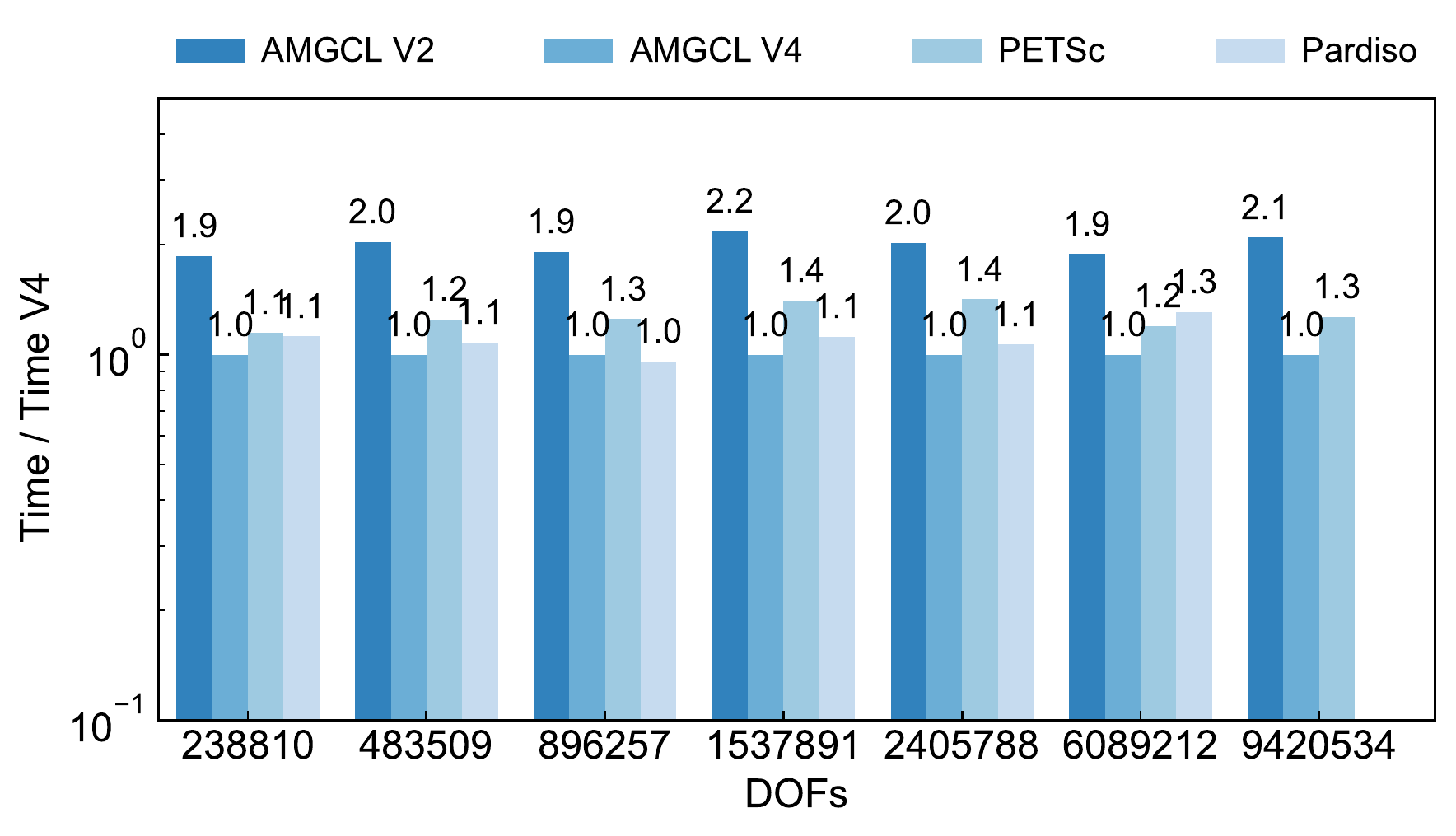}
    \includegraphics[width=0.48\linewidth]{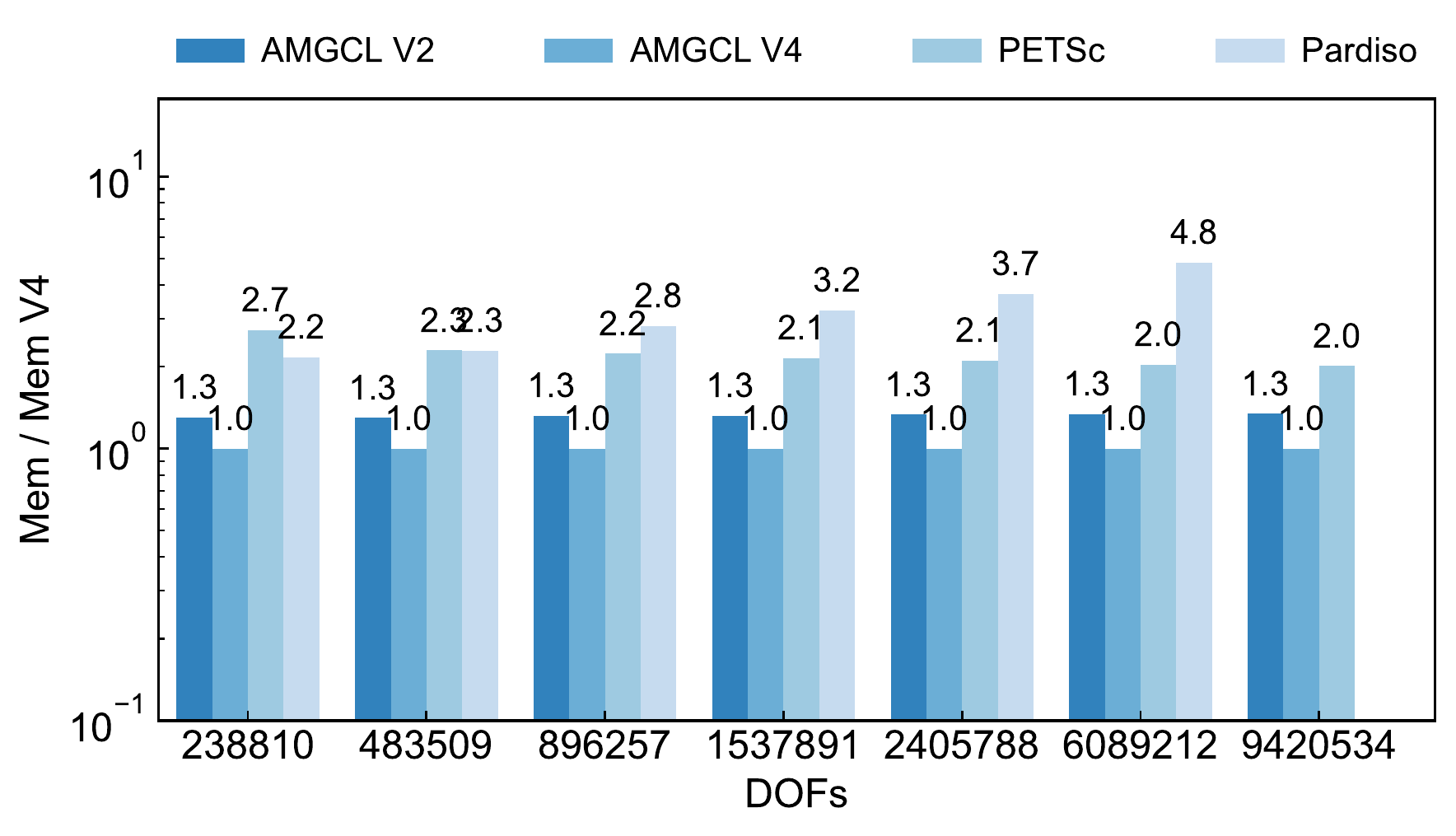}
    \caption{Solution time and memory usage for the sphere packing
    problem.}
    \label{figure:Case3_Performance}
\end{figure}

The final case we consider is a complex sphere packing flow problem with
non-uniform cell size distribution and large cell size contrast
(\Cref{figure:Case3_mesh}). A box-shaped domain of $85 \times 85 \times 85
\si{\micro\meter}$ is used where the 21 spheres with a radius of
$22\si{\micro\meter}$ are packed randomly with the resulting porosity of
36.4\%.  Similar to the converging-diverging tube problem, a pressure drop of
$1\si{\pascal}$ is applied on both ends of the domain in z direction
($z=0\si{\micro\meter}$ and $z=85\si{\micro\meter}$). The no-slip wall boundary
conditions of \lang{$\mathbf{u}=\left(0,0,0\right)^T$} are imposed on all sphere
surfaces and box surfaces in Y and X directions. The 7 refinement mesh
levels are generated using Gmsh. \lang{The mesh and the computed velocity field
are shown in~\Cref{figure:Case3_mesh}.} \Cref{table:SpherePacking} shows the
problem size at each level\lang{, and} \Cref{figure:Case3_Performance} presents the
solution time and the memory usage scaled by the CSR matrix size and by the
AMGCL V4 solver results.

\begin{table}
\caption{Linear system size (DOFs), number of non-zeros (NNZs), and
    linear system CSR storage memory for the sphere packing problem.}
\centering
\begin{tabular}{rrrrrr}
\hline
Cells   & DOFs      & NNZs        & CSR Memory \\ \hline
9~487   & 238~810   & 5~021~060   & 176.87 Mb  \\
19~373  & 483~509   & 10~400~000  & 397.01 Mb  \\
36~430  & 896~257   & 20~200~000  & 687.07 Mb  \\
63~382  & 1~537~891 & 36~100~000  & 1.18 Gb    \\
100~255 & 2~405~788 & 58~400~000  & 1.87 Gb    \\
259~009 & 6~089~212 & 157~000~000 & 4.90 Gb    \\
404~019 & 9~420~534 & 248~874~496 & 7.96 Gb    \\
\hline
\end{tabular}
\label{table:SpherePacking}
\end{table}

The matrix structure in this case is favorable to the
direct Pardiso solver, since the solver shows the relatively low memory usage of approximately 10 times the
size of the CSR matrix, as opposed to 21 and 27 times for the unit cube and the
converging-diverging tube problems at the largest problem sizes. This could
possibly be explained by the significant amount of tetrahedron faces with the
Dirichlet boundaries modeling the no-slip conditions at the sphere walls. As
a result, the direct Pardiso solver shows almost linear scalability and only
starts to slow down at the largest problem size it can handle on our machine.
Up to that point, it shows the solution times comparable with AMGCL V4, which
shows the best solution times among the iterative solvers for this problem.
However, the high memory requirements still remain an issue for the Pardiso
solver.

The simplest AMGCL V1 version with the monolithic preconditioner shows worse
than linear scalability, but still is able to outperform V2 even for the
largest problem. Its efficiency is explained by the fact that
AMGCL V1 makes twice fewer iterations for the sphere packing problem than in the
cases of the unit cube and converging-diverging tube problems, which probably has
the same reasons the Pardiso solver is working so well here.

Among the rest of the iterative solvers, AMGCL V2 shows the worst
performance throughout the whole range of the problem sizes, while AMGCL V4
consistently has the best solution times and is about 2 times faster than V2
and 10--30\% faster than PETSc. Memory-wise, AMGCL solvers are still the most
efficient. The V4 version requires about twice less memory than the PETSc
solver for all problem sizes starting with the $0.4\times10^6$~DOFs one. As in
the previous cases, the constant memory overhead inherent to PETSc skews the
ratio for the smallest problem size.

\section{Discussion}

We have analyzed the efficiency of several solvers for large sparse linear
systems obtained from discretization of the Stokes equations with a discontinuous
Galerkin finite element method. As expected, the direct Pardiso solver from MKL
demonstrated the worst scalability in terms of both the solution time and the
memory footprint of the solution. A Krylov subspace iterative method
preconditioned with a monolithic preconditioner ignoring the system
structure scales linearly with the system size with respect to
memory, but the solution time has faster than linear growth.

The solvers that use the Schur pressure correction preconditioner
accounting for the problem structure show equally good scalability.
However, the PETSc version is 2--4 times faster than the AMGCL V2 version, even
though both versions work with scalar matrices and use a full precision approach.
AMGCL is able to outperform PETSc for two out of three benchmarks by up to 40\%
with versions V3 (using the block-valued backend for the flow subsystem), and
V4 (using both the block-valued backend and mixed precision approach). This
means that if the performance of the baseline AMGCL V2 could be improved to be on par with the
PETSc implementation, then the performance of the AMGCL V4 solver
could be considerably faster than the PETSc version. This, however, is out of
scope of the current work.

Comparing the baseline AMGCL V2 performance to V3 shows that the switch from
the scalar to the block value type reduces the setup complexity by 85\%, the
memory footprint of the method by about 20\% and achieves the total speedup of
30\% to 40\%.  Further, using a mixed precision approach (single
precision preconditioner with double precision iterative solver) the memory
footprint of the method is further reduced by 30\% and the total solution time
is decreased by another 30\%. Overall, the V4 solver requires
approximately 40\% less memory and is 2 to 4 times faster than the V2 solver,
even though the number of iterations between versions V2~--~V4 of the solver
stays practically constant. In other words, as long as using the single
precision preconditioning does not affect the convergence of the solution, the
improvements described here should not change the algorithmic scalability of
the method, but should simply accelerate the underlying operations.

It should be noted that AMGCL versions from V2 to V4 are using the
same code. By using the \CXX metaprogramming techniques such as policy-based
design, partial template specialization, and free functions in the library
code, we are able to select the most efficient implementation simply by
changing the backend definition in the solver template parameters.  Of course,
the \CXX metaprogramming does not guarantee high performance by itself, and the
same approach could be implemented in a pure C library, but it would require
much bigger effort from the library developers. The results in this work were
presented for the new DG FEM method, but we believe the same techniques could
be used to accelerate the solution of any Stokes-like system.

\section{Acknowledgements}

Authors would like to thank Dr. Karsten Thompson (Louisiana State University),
Dr. Richard Hughes (Louisiana State University), and Dr. Xiaozhe Hu (Tufts
University) for providing support and helpful suggestions.

\bibliographystyle{elsarticle-num}
\bibliography{ref}

\end{document}